\documentclass[aps,amssymb,amsfonts,amsmath,a4paper,preprint]{revtex4}
\usepackage{longtable}

\newif\ifpdf
\ifx\pdfoutput\undefined
  \pdffalse
\else
  \pdfoutput-1
  \pdftrue
\fi
\ifpdf
     \usepackage[pdftex]{graphicx}
     \DeclareGraphicsExtensions{.pdf}
\else
     \usepackage[dvips]{graphicx}
     \DeclareGraphicsExtensions{.eps}
\fi

\makeatletter
\renewcommand{\@seccntformat}[1]{\large \csname the#1\endcsname .
\hspace{0.5em}}
\makeatother
\numberwithin{equation}{section}

\begin{document}

\title{$3D$ chaotic model for sub-grid turbulent dispersion in 
 Large Eddy Simulations}

\author{Guglielmo Lacorata$^1$, Andrea Mazzino$^2$ and Umberto Rizza$^1$ 
}

\address{$^1$CNR, Institute for Atmospheric and Climate Sciences, 
str. Lecce-Monteroni, 73100 Lecce, Italy
}
\address{$^2$Department of Physics, University of Genova, and INFN, CNISM, 
Section of Genova,   via Dodecaneso 33, 16146 Genova, Italy}

\maketitle

\newpage

\centerline{Abstract}
{\noindent 
We introduce a $3D$ multiscale kinematic velocity field as a model to simulate 
Lagrangian turbulent dispersion. The incompressible velocity field is a 
nonlinear deterministic function, periodic in space and time, that generates chaotic 
mixing of Lagrangian trajectories. Relative dispersion properties, e.g. the Richardson's law, 
are correctly reproduced under two basic conditions: 1) the velocity amplitudes   
of the spatial modes must be related to the corresponding wavelengths through the 
Kolmogorov scaling; 2) the problem of the lack of "sweeping effect" of the small eddies by the large  
eddies, common to kinematic simulations, 
has to be taken into account. We show that, as far as Lagrangian dispersion is concerned, 
our model can be successfully applied as additional sub-grid contribution for Large Eddy Simulations 
of the planetary boundary layer flow.      
}

\newpage

\section{Introduction}
Lagrangian transport and mixing of trajectories in turbulent flows,
e.g. the planetary boundary layer (PBL), or even the ocean mixed layer
(OML), can be studied through models known as Large Eddy Simulations, 
 or, briefly, LES (Lilly, 1967, Leonard, 1974, Moeng, 1984). 
 According to the LES strategy, only the large-scale motion
associated to the largest turbulent eddies is explicitly solved, while
the small-scale dynamics, partly belonging to the inertial range of
scales, is described in a statistical consistent way (i.e., it is
parameterized in terms of the resolved, large-scale velocity and
temperature fields). \\
It is commonly believed, and actually shown by means of many numerical
experiments, that the effect of the small-scale parameterized eddies
does not considerably affect the large-scale explicitly resolved motion.
In view of this fact, the LES strategy appears 
suitable to describe large-scale properties as, in way of example, 
trajectory dispersion driven by the resolved 
velocity modes.
 The finite spatial resolution of any LES, obviously,
implies the lack of dynamical informations on the small-scale advecting 
velocity necessary to properly describe particle trajectories.
If we want to take into account also the small-scale (unresolved) 
contribution to Lagrangian motion, we need a model for replacing the
sub-grid components that are filtered out. 
This point assumes particular importance if 
 one is interested in pre-asymptotic dispersion of a cloud of tracer, e.g. over spatio-temporal scales 
 comparable to the characteristic spatio-temporal scales of a LES domain. Asymptotic eddy-diffusion 
 is, indeed, unaffected from the small-scale details of the dynamics. 
 Turbulent-like motions of particles 
can be generated  
 by means of either stochastic models of dispersion (Thomson, 1987) or kinematic models like, e.g.,  
 a series of unsteady random Fourier modes (Fung et al., 1992, Fung and Vassilicos, 1998).  
 Our aim here is to exploit the possibility of a fully deterministic nonlinear dynamical system to reproduce 
 the same Lagrangian particle dispersion properties as observed in actual turbulent 
 flows.  \\  
In this respect,  
we introduce and analyse a multiscale $3D$, incompressible, kinematic velocity field that 
generates chaotic Lagrangian trajectories, and show how the  
problem of the lack of "sweeping effect" of the small eddies by the large eddies (Thomson and Devenish, 2005), 
can be eliminated, from a Lagrangian point of view, 
by an appropriate redefinition of the spatial coordinates, such that the mean square 
particle displacement may follow the expected $t^3$ Richardson's law.  
 The kinematic model has zero mean field,  
 so, once it is employed as sub-grid model in LES, 
 the absolute dispersion properties of generic particle distributions advected by 
 the large scale eddies remain unchanged, except for particular cases like, for example, 
 particle sources near the ground, where a large fraction of the energy is contained in the subgrid scales. 
 As far as relative transport properties are concerned, instead, we show that it is possible, 
 in some sense, to extend 
 the turbulent relative dispersion law of the resolved large scales down to the unresolved small scales, 
 by means of the kinematic field.     
 
The paper is organized as follows: 
 in section \ref{sec:fsle} we recall the scale-dependent characteristics of 
 relative dispersion; 
in section  \ref{sec:kinmod} we introduce our kinematic model;  
 section \ref{sec:les} contains a description of the LES model; 
in section  \ref{sec:results} we report the results obtained from our analysis and, last, 
in section \ref{sec:conclusions} we discuss what conclusions can be drawn after this work 
and possible perspectives.

\section{Main aspects of Lagrangian dispersion}
\label{sec:fsle}

Let $\vec{r}=(x,y,z)$ and $\vec{V}=(u,v,w)$ be the position vector and the velocity vector, respectively, 
of a fluid particle. We define   
$\Delta{\vec r}(t)$ as the distance between two particles at time $t$ and 
$\Delta{\vec V}(\delta)$ as the velocity difference between two particles at distance 
$|\Delta{\vec r}|=\delta$. 
 The two-particle statistics we consider as diagnostics of relative dispersion are the following:
\begin{itemize}
	\item the finite-scale Lyapunov exponent $\lambda(\delta)$ (FSLE), defined as the 
	inverse of the mean time $\langle \tau_e \rangle$ taken by the particle separation 
	to grow from $\delta$ to $\beta \cdot \delta$, with $\beta > 1$, multiplied by ${\rm ln}(\beta)$: 
	$\lambda(\delta)= {\rm ln} (\beta) / \langle \tau_e \rangle$, for any $\delta$ and 
	$\beta \sim O(1)$ (Artale et al., 1997, Boffetta et al., 2000);   
	\item the classic mean square particle separation $R^2(t)$ as function of time $t$, 
	$R^2(t) = \langle |\Delta{\vec r}(t)|^2 \rangle$ averaged over 
	an ensemble of trajectory pairs.
\end{itemize}
As far as absolute dispersion is concerned, we define the two-time, one-particle statistics: 
\begin{itemize}
	\item $S^2(t)=\langle [{\vec r}(t)-{\vec r}(0)]^2\rangle-\langle[{\vec r}(t)-{\vec r}(0)]\rangle^2$, 
	averaged over an ensemble of trajectories. In case of zero mean advection, the $\langle[{\vec r}(t)-{\vec r}(0)]\rangle$ 
	term of course vanishes.  
\end{itemize}

We want to stress here  
that the relative mean square displacement, $R^2(t)$ is not fully equivalent, for 
physical reasons we will briefly discuss below, to the fixed-scale analysis based on the FSLE.         
The FSLE is an exit-time technique and, in a Lagrangian context, it  
	may be also referred to as finite-scale dispersion rate. 
	This quantity was formerly introduced in the framework of dynamical systems theory (Aurell et al., 1996, 1997) and 
later exploited for treating finite-scale Lagrangian relative dispersion as a finite-error predictability problem 
 (Lacorata et al. 2001, Iudicone et al., 2002, Joseph and Legras, 2002, LaCasce and Ohlmann, 2003). 
 The physical reason why the FSLE is to be preferred, as analysis technique of the relative dispersion process, with respect to the time dependent mean square displacement, is the following: 
 particle separation changes behavior in correspondence of certain characteristic lengths of the velocity field, not 
 much in correspondence of certain characteristic times; reasonably, particle pairs are not supposed to enter a certain 
 dispersion regime (e.g. one of those discussed below) at the same time; the arrival time of the particle separation 
 to a certain threshold (i.e. the exit time from a certain dispersion regime) can be usually subject to strong fluctuations. So, if one computes a fixed time average of the particle separation risks to have in return misleading information, because 
 of overlap effects between different dispersion regimes; 
 if one considers a fixed scale average of the dispersion rate, 
 instead, the relative dispersion process, as function of the particle separation scale, is described in more consistent physical terms, as already widely established in previous works, see Boffetta et al. (2000) for a review.  
  
We will describe now, briefly, three major relative dispersion regimes that generally occur.   

Chaos is a common manifestation of nonlinear dynamics and it implies exponential 
separation of arbitrarily close trajectories (Lichtenberg and Lieberman, 1982), i.e. 
sensitivity to infinitesimal errors on the initial conditions (Lorenz, 1963). 
The mean growth rate $\lambda$ is known as maximum Lyapunov exponent (MLE): 
\begin{equation}
|\Delta{\vec r}(t)| \sim 	|\Delta{\vec r}(0)| {\rm e}^{\lambda t}
\label{eq:LLE}
\end{equation}
If the trajectories refer to Lagrangian particles, as is the case in this work, 
$\lambda$ can be also called Lagrangian Lyapunov exponent (LLE). 
The regime (\ref{eq:LLE}) lasts  
as long as the particle separation remains {\it infinitesimal} relatively to the  
characteristic lengths of the velocity field. In the limit $\delta \to 0$  
 the velocity field is considered {\it smooth}: $|\Delta{\vec V}(\delta)| \sim \delta$;  
 the dispersion rate is, therefore, independent from the separation scale:   
 $\lambda(\delta) = \lambda(0) = \lambda$, i.e. the FSLE is equal to the LLE.   
Even a regular, in other terms non turbulent, velocity field may generate Lagrangian 
chaos (Ottino, 1989) provided that the velocity field is, of course, nonlinear and, generally, time-dependent. 

Standard diffusion (Taylor, 1921) means, in a few words, Gaussian distribution of the particle separation, with 
zero mean and variance linearly growing in time, an asymptotic regime 
occurring after the full decay of the correlations between the particle velocities:
\begin{equation}
R^2(t) = 4 \, D_E \,t
\label{eq:Diff}	
\end{equation}
The quantity $D_E$ is known as eddy diffusion coefficient (EDC) and it represents the 
effective diffusivity of trajectories caused by the largest eddies in a multi-scale 
structured flow (Richardson, 1926). The $4$ factor (instead of $2$) in eq. (\ref{eq:Diff}) appears  
since we are considering relative (and not absolute) dispersion.  
By dimensional argument, it can be shown that the dispersion rate 
 must scale with the particle separation as $\lambda(\delta) \sim \delta^{-2}$.  
This behavior may be observed only if 
the spatial domain is much larger than the Lagrangian correlation scale which is, typically, 
of the order of the eddy maximum size. In other words, only when the distance between two particles is 
sufficiently larger than the correlation length of the velocity field,  
the relative velocity can be approximated by a stochastic process with zero mean and 
finite correlation time, which implies long time standard diffusive behavior.      

In fully developed turbulence, 
between the two regimes described above, chaos and diffusion, 
there exists an intermediate regime inside the so called inertial range of scales,   
characterized by a direct energy cascade from large to small vortices (see Frisch, 1995, for a review), 
with mean energy flux $\epsilon$. Inside the inertial range, 
relative dispersion follows a super-diffusive scaling with time, 
according to the power law empirically discovered by Richardson (1926): 
\begin{equation}
R^2(t) = C_R \,\epsilon \,t^3
\label{eq:Richardson}		
\end{equation}
where $C_R$ is known as the non dimensional Richardson's constant. 
Recent experimental and numerical studies 
agree about a value of the Richardson's constant $C_R \simeq 5 \cdot 10^{-1}$ 
(Ott and Mann, 2002, Boffetta and Sokolov, 2002, Gioia et al., 2004).   
The Richardson's 
law (\ref{eq:Richardson}) can be derived from the fundamental assumption  
 of the theory of turbulence stating that $|\Delta{\vec V}(\delta)|^2 \sim \delta^{2/3}$ inside the inertial range 
 (Frisch, 1995). 
 It can be verified, by a simple dimensional argument, that the equivalent of 
 (\ref{eq:Richardson}) in terms of FSLE is $\lambda(\delta) = \alpha \delta^{-2/3}$, 
 where $\alpha^3$ is a quantity of the same order as $\epsilon$ (Gioia et al., 2004, Lacorata et al., 
 2004).  

In the next section we introduce the $3D$ kinematic model and discuss its main characteristics.

\section{The $3D$ kinematic model}
\label{sec:kinmod}

\begin{figure}[h]
	\begin{center}
		\includegraphics[angle=0,width=15cm,height=15cm]{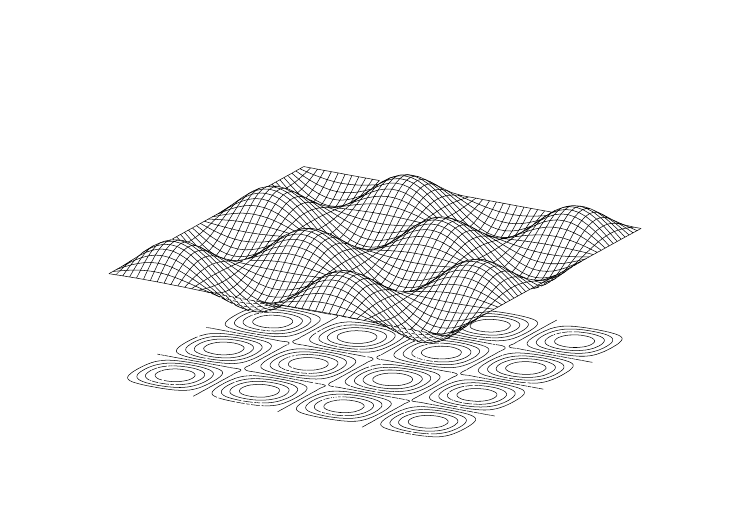}
	\end{center}
	\caption{}
\label{fig:psi_km}
\end{figure}

The time evolution of a fluid particle position 
$\vec{r} =(x,y,z)$, given a velocity field $\vec{V} =(u,v,w)$, is the solution of: 
\begin{equation}	
\frac{d\vec{r}}{dt}(t) = \vec{V}(\vec{r},t)
\label{eq:lagrange}
\end{equation}
For a fixed initial condition, $\vec{r}(0) =(x(0),y(0),z(0))$, we assume there is  
one and only solution to eq. (\ref{eq:lagrange}). A nonlinear velocity field,   
 as pointed out in the previous section, even very simple, is necessary for  
 having exponential growth of arbitrarily small   
 errors on the initial conditions. In analogy with $2D$ cellular flows defined in terms 
 of one (time dependent) stream-function (Solomon and Gollub, 1988, Crisanti et al., 1991), we 
 define two (time dependent) stream-functions, $\Psi_I$ and $\Psi_{II}$, as follows:    
 \begin{eqnarray}
	\Psi_{I}(y,z,t) &=& (A/k_3) \cdot sin(k_2(y-\xi_2 sin(\omega_2 t))) 
	\cdot sin(k_3(z-\xi_3 sin(\omega_3 t)) \\
	\Psi_{II}(x,z,t) &=& (A/k_3) \cdot sin(k_1(x-\xi_1 sin(\omega_1 t))) 
	\cdot sin(k_3(z-\xi_3 sin(\omega_3 t))) 
	\label{eq:dsf}
\end{eqnarray}
where: $A$ is the velocity scale; $\vec{k}=(k_1,k_2,k_3)$ is the wavevector corresponding to   
the wavelengths $(l_1,l_2,l_3)$ of the flow according to the usual relations $k_i=2\pi/l_i$, 
for $i=1,2,3$;   
$\vec{\xi}=(\xi_1,\xi_2,\xi_3)$ is the amplitude vector and 
$\vec{\omega}=(\omega_1,\omega_2,\omega_3)$ is the   
pulsation vector of the time-dependent perturbative terms.  
If we formally associate the two stream-functions $\Psi_{I}$ and $\Psi_{II}$ to the 
components of a `potential vector' 
$\vec{\Psi}=\{\Psi_{I}(y,z,t), \Psi_{II}(x,z,t), 0\}$, we can define a 
$3D$, non divergent, velocity field as  
$\vec{V}=-\vec{\nabla} \times \vec{\Psi}$. On the basis of this definition, 
the three components of the velocity field are:
\begin{eqnarray}
	u &=& \frac{\partial \Psi_{II}}{\partial z}\\
	v &=& -\frac{\partial \Psi_{I}}{\partial z}\\
	w &=& -\frac{\partial \Psi_{II}}{\partial x}+\frac{\partial \Psi_{I}}{\partial y}
	\label{eq:vel_km}
\end{eqnarray}
We name the kinematic velocity field (III.4), (III.5), (III.6) as 
the (one-mode) double stream-function (DSF) field. The explicit expressions of the three 
velocity components results to be: 
\begin{eqnarray}
	u &=& A sin(k_1(x-\xi_1 sin(\omega_1 t))) cos(k_3(z-\xi_3 sin(\omega_3 t))) \\
	v &=& - A sin(k_2(y-\xi_2 sin(\omega_2 t))) cos(k_3(z-\xi_3 sin(\omega_3 t)))  \\
	w &=& - A \frac{k_1}{k_3} cos(k_1(x-\xi_1 sin(\omega_1 t))) sin(k_3(z-\xi_3 sin(\omega_3 t)))\nonumber  \\ 
	  &+& A \frac{k_2}{k_3} cos(k_2(y-\xi_2 sin(\omega_2 t))) sin(k_3(z-\xi_3 sin(\omega_3 t)))
	\label{eq:vel3_km}
\end{eqnarray} 
By setting $k_1=k_2=k$, $k_3=2k$, together with   
$\xi_1=\xi_2=\xi_3$, and $\omega_1\approx\omega_2\approx\omega_3$, it can be shown that  
the chaotic Lagrangian motion, generated by the DSF field is, on average, isotropic to a good extent, as 
discussed below relatively to the multiscale version of the DSF model. 
At this regard, we would like to stress that Lagrangian chaos has the worth-noting 
advantage to simulate an almost isotropic trajectory dispersion even in a not exactly isotropic velocity field. 
The one-mode DSF model (III.7), (III.8) and (III.9)
is characterized by a periodic pattern of $3D$ quasi-steady eddies of 
size $\sim l_1$, with typical convective velocity $\sim A$ and turnover time defined as $\tau = l_1 / A$.   
The two stream-functions $\Psi_I$ and $\Psi_{II}$, if taken singularly, describe   
$2D$ convective velocity fields (Solomon and Gollub, 1988)  
as illustrated in Fig. \ref{fig:psi_km} for the steady case. 
There is no unique way, 
of course, to get a $3D$ generalization of $2D$ cellular fields. 
The DSF field is one of the simplest options.    
  
The extent of the chaotic layer, i.e. the region of the space where initially 
close trajectories move apart from each other exponentially fast in time, 
  depends on the working point $(\xi_1,\xi_2,\xi_3,\omega_1,\omega_2,\omega_3)$ in the 
 perturbative parameter space (Chirikov, 1979). It can be numerically proved that 
 a good "efficiency" of chaos, as mechanism of 
  trajectory mixing all over the space, is obtained by setting the perturbation periods 
 to the same order as the turnover time, 
 and the perturbation amplitudes to a fraction of the cell size (Crisanti et al., 1991). 
 We adopt the following definitions (valid for both the one-mode and the multi-mode DSF model):  
 $\omega_1 = 2 \pi / \tau^{-1}$, $\omega_2 = \sqrt[3]{2} \omega_1$, $\omega_3 = (\pi / 3) \omega_2$, 
 and $\xi_i/l_i = 0.25$, for $i=1,2,3$. Although the three perturbation frequencies are of the same order, 
 they have not exactly the same numerical value. The ratios between them are, in fact, "irrational" 
 numbers (within the obvious limits imposed by the computer finite precision) of order $O(1)$.   
 This precaution is adopted in order to avoid virtually possible (even though highly improbable) 
 "trapping" effects of a particle inside a convective cell due to some unwanted peculiar phase coincidence.   
 

As long as only one characteristic scale is involved, 
the Lagrangian dispersion properties of the DSF model can be described by the following 
two regimes:    
 
 Lagrangian chaos, $|\Delta{\vec r}(t)| \sim |\Delta{\vec r}(0)| \, {\rm e}^{\lambda t}$, 
 as long as $|\Delta{\vec r}(t)| \ll l_1$, 
 with the LLE  
 of the order of the inverse turnover time, $\lambda \sim \tau^{-1}$; 
 
 standard diffusion,  $|\Delta{\vec r}(t)|^2 \sim 4 \, D_E \, t$, 
 when $|\Delta{\vec r}(t)| \gg l_1$, with the EDC $D_E \sim l_1 \cdot A$.  
 
\begin{figure}[h]
	\begin{center}
		\includegraphics[angle=0,width=15cm,height=15cm]{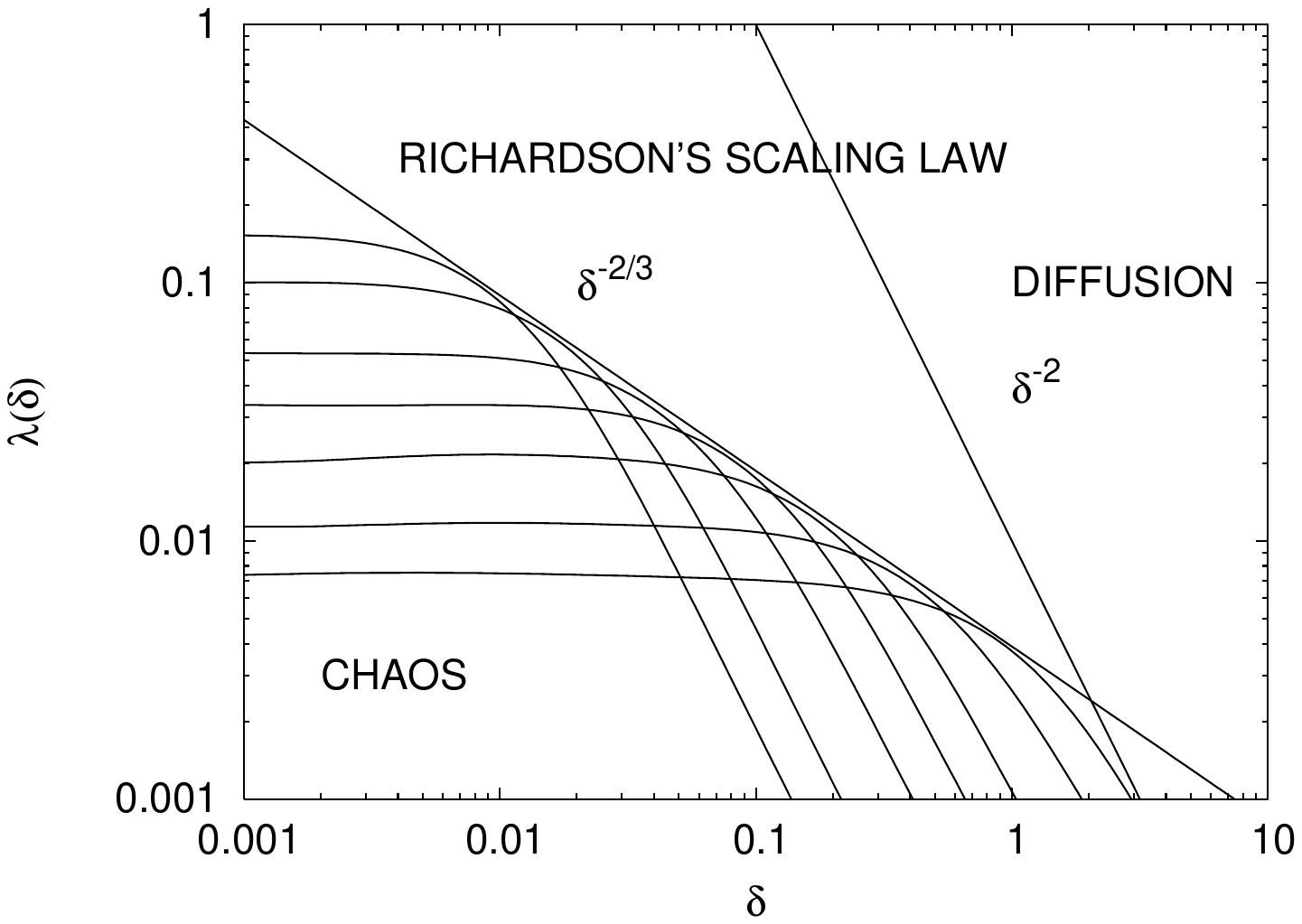}
	\end{center}
	\caption{}
\label{fig:regimes_km}
\end{figure}

 We can see in Fig. \ref{fig:regimes_km} the FSLE of the one-mode DSF model 
 at various spatial wavelengths $l_1$ with the velocity amplitude scaling as 
 $A \sim l_1^{1/3}$. The FSLE is computed over a range of scales $\delta_1,...,\delta_N$ 
 such that $\delta_{n+1}=\beta \cdot \delta_n$, for $n=1,N$, with $\beta=\sqrt{2}$, and 
 $\delta_N \gg l_1$.  
 The initial particle separation, in each case, is set much smaller 
 than the eddy size, $\delta_1 \ll l_1$, 
 and the integration time step of the numerical simulations is 
 set much smaller than the turnover time scale, $dt \ll \tau$. 
 At very small and very large particle separation, the two regimes described above,  
chaos and diffusion, correspond to the $\lambda(\delta)=const.$ and 
$\lambda(\delta) \sim \delta^{-2}$ laws, respectively. As consequence of the 
$A \sim l_1^{1/3}$ scaling, the horizontal levels (i.e. the LLE $\lambda$) 
scale as $\tau^{-1} \sim l_1^{-2/3}$, as seen by the alignment of the FSLE "knees" 
along the $\sim \delta^{-2/3}$ law. At this stage, there is 
no need to take into account the problem of the "sweeping effect", which 
will make its appearance in the multiscale case. 
The DSF model can describe, indeed, velocity fields with a series of 
 spatial modes:
\begin{eqnarray}
	\Psi_I(y,z,t)&=& \sum_n \Psi_I^{(n)}(y,z,t)\\
	\Psi_{II}(x,z,t)&=& \sum_n\Psi_{II}^{(n)}(x,z,t)
\end{eqnarray}
for $n=1,N_m$ where $N_m$ is the number of modes. 
The $n$-th term in the sums is characterized by the set 
$(A^{(n)},\vec{k}^{(n)},\vec{\xi}^{(n)},\vec{\omega}^{(n)})$ representing 
velocities, wavenumbers, oscillation amplitudes and pulsations, respectively, 
 of the $n$-th mode. The eddy turnover times are defined as $\tau^{(n)}=l_1^{(n)} / A^{(n)}$. 
 Let us assume that, for each mode $n=1,N_m$,    
$k_1^{(n)}= k_2^{(n)}$, and $k_3^{(n)}= 2k_1^{(n)}$, and that 
$k_1^{(n)}= k_1^{(1)}\cdot\rho^{n-1}$, with $\rho > 1$.
We model the turbulent relative dispersion by assigning the Kolmogorov scaling  
to the velocity as function of the wavevector amplitude:
\begin{equation}
A^{(n)} = \sqrt{2C_K} \left(\frac{\epsilon}{k_1^{(n)}}\right)^{1/3}
\label{eq:scaling}
\end{equation}
where $C_K$ is the equivalent Kolmogorov constant and $\epsilon$ is the equivalent 
mean energy flux from large to small scales inside the inertial range of a turbulent flow  
(even though, of course, no energy cascade occurs in kinematic fields).  
Some considerations based on the geometry of the flow show that, given $N_m$ modes, 
the effective inertial range of the field corresponds to the interval 
$\left[k_{max},k_{min}\right] \simeq \left[2k_1^{(N_m)},4k_1^{(1)}\right]$. This fact is due 
to the spatial structure of the three-dimensional convective cells, in particular it can be shown 
that each wavelength includes two (dynamically equivalent) adjacent cells of half wavelength edge. So that, 
the largest correlation length between two particles results to be nearly $1/2$ the edge of 
the largest cells (i.e. about $1/4$ the largest wavelength), and turns out to be the actual upper bound 
of the inertial range; the smallest 
 cell edge is $1/2$ the smallest wavelength, and 
 turns out to be the actual lower bound of the inertial range, as confirmed by the numerical simulations. 
 The constant $C_K \sim 10^{-1}$ determines the order of the equivalent Richardson's constant 
 $C_R \sim 10^{-1}$ of the kinematic simulation. For instance, it can be verified that a value 
 $C_K=0.25$ corresponds to having $C_R \simeq 0.5$ for any energy flux $\epsilon$. Eventually, we will see   
 that $C_K$ is the free parameter to adjust for the fine tuning of the DSF field to the LES field. 
 
Recently, some authors have raised the question if a kinematic velocity field, made of a series 
of fixed eddies of various length scales, even though subject to periodic oscillations  
around their mean location as occurs in the DSF field, 
can really reproduce the right scaling law of the relative dispersion as predicted by Richardson. 
Thomson and Devenish (2005) have shown that, even though for each mode the velocity amplitude is related to the 
spatial wavelength through the Kolmogorov scaling (\ref{eq:scaling}), the lack of advection of the small eddies 
by the large eddies, as is the case for kinematic simulations, can modify the behavior of the 
mean square relative displacement inside the inertial range. 
In particular, if the integration time step becomes sufficiently small, 
i.e. basically $dt < l_{min}/v_{max}$, where $l_{min}$ and $v_{max}$ are the smallest vortex length and the maximum advecting velocity in one point, respectively, 
relative dispersion is found to scale as $R^2(t) \sim t^{\gamma}$ with $\gamma > 3$. 
 In the two limit cases, as discussed in Thomson and Devenish (2005), of zero mean field and strong mean field, 
 the exponent of the scaling law turns out to be, respectively, $\gamma=9/2$ in one case and $\gamma=6$ in the other. 
 
 This problem may be overtaken by considering the kinematic model as a two-particle dispersion model, computed in the 
 reference frame of the mass center of the particle pair. The technique consists in replacing the absolute coordinates 
  that appear in the arguments of the DSF sinusoidal functions with relative coordinates. 
  If at time $t$  
  two particles have coordinates $(x_1(t),y_1(t),z_1(t))$ and $(x_2(t),y_2(t),z_2(t))$, we redefine, 
  for every time $t$,  $x_i(t) \to x_i(t)-x_M(t)$, 
  $y_i(t) \to y_i(t)-y_M(t)$ and $z_i(t) \to z_i(t)-z_M(t)$, for $i=1,2$,  
  where $x_M(t)=(x_1(t)+x_2(t))/2$, $y_M(t)=(y_1(t)+y_2(t))/2$ and $z_M(t)=(z_1(t)+z_2(t))/2$ are the mass center coordinates of the two particles at time $t$. This means that each particle pair moves in its own kinematic field 
  anchored to its mass center, and is therefore subject to the relative dispersion caused by eddies that are 
  advected together with the particles by the large scale velocity field.  
  There is no relative advection between eddies of different size, in the sense that all the convective structures 
  are advected at the same speed, but, at least, the "fast crossing" of a particle pair 
  through the convective cell pattern is, in this way, eliminated. This is confirmed by the numerical simulations presented below. In absence of an additional large scale field, we can leave the first 
  mode of the model unchanged by the relative coordinates technique, that is the coordinates appearing in the 
  arguments of the $n=1$ term are absolute coordinates $(x(t),y(t),z(t))$. This assures the global spatial 
  averaging of the dispersion process. The whole procedure in terms of equations can be written as:
\begin{eqnarray}
\frac{dx}{dt} &=& 
\frac{\partial}{\partial z}\left\{\Psi_{II}^{(1)}(x,z,t) + \sum_{n=2}^{N_m} \Psi_{II}^{(n)}(x_R,z_R,t)\right\} \\
\frac{dy}{dt} &=& 
-\frac{\partial}{\partial z}\left\{\Psi_{I}^{(1)}(y,z,t) + \sum_{n=2}^{N_m} \Psi_{I}^{(n)}(y_R,z_R,t)\right\} \\
\frac{dz}{dt} &=& 
-\frac{\partial}{\partial x}\left\{\Psi_{II}^{(1)}(x,z,t) + 
\sum_{n=2}^{N_m} \Psi_{II}^{(n)}(x_R,z_R,t)\right\} \nonumber \\
 &+& \frac{\partial}{\partial y}\left\{\Psi_{I}^{(1)}(y,z,t) + 
 \sum_{n=2}^{N_m} \Psi_{I}^{(n)}(y_R,z_R,t)\right\}
\end{eqnarray}
where $(x,y,z)$ are the absolute coordinates of one of the two particles and $x_R,y_R,z_R$ are the relative coordinates with respect to the mass center of the particle pair. 
Equations (III.13), (III.14) and (III.15) define the multi-scale DSF model.  
 The DSF velocity field has the structure of a periodic pattern  
 of $3D$ non steady convective cells, of size varying within a given range of scales, 
 fixed in space but subject to 
 periodic oscillations around their equilibrium positions.  
 Despite the eddies have infinite lifetime, i.e. the one-point Eulerian correlations do not decay,  
turbulent-like trajectories can be generated even from a non turbulent 
velocity field, under certain conditions, by means of the effects of Lagrangian chaos acting at every 
scale of motion. 

In the next section we describe the LES  
experiment that provide the large-scale velocity field of a convective planetary boundary layer, and the way 
 the DSF velocity field is coupled to the LES as subgrid kinematic model. At this regard, we observe that, 
 in presence of the large scale flow provided by the LES, 
 the first spatial mode of the multi-scale DSF model needs no longer 
 to be treated differently from all the other modes, since a large scale mixing is assumed to occur anyway 
 favoured by the turbulent velocity field of the LES. In any case, we will not modify the definition of the 
 multi-scale DSF model, as established by (III.13), (III.14) and (III.15), even when nested in the LES. 

\section{Numerical experiments}
\label{sec:les}

\subsection{The LES velocity field}
\label{sec:exp}

The LES model 
advances in time the filtered equations for the temperature
and the velocity field, coupled via the Boussinesq approximation. 
Subgrid scale momentum and heat eddy-coefficients are expressed 
in terms of the
subgrid turbulent kinetic energy, the evolution equation of which
is integrated by the LES model. 
The numerical simulations have been performed on a
$128^3$ cubic lattice, biperiodic in the horizontal plane. The LES code is
pseudospectral in the horizontal plane, while it is discretized with finite
differences in the vertical direction. \\
Such a model has been widely used and tested to
investigate basic research problems in the framework of boundary layer
flows (see, for example, Moeng and Wyngaard, 1988;
Moeng and Sullivan, 1994; Porte Agel et al. 2000; 
Antonelli et al., 2003; Gioia et al. 2004, Rizza et al. 2006, among the
others). The major reference papers for the LES model we use  
are Moeng (1984), and Sullivan et al. (1994).

In the present study, we have performed one numerical experiment
characterized by a stability
parameter $z_i/L_{mo} \simeq 15$, where $z_i$ is the mixing layer height
and $L_{mo}$ is the Monin-Obukhov length, provides a measure of the
atmospheric stability.  According to Deardoff (1972), the convective
regime settles in if $z_i/L_{mo} > 4.5$.  
The characteristic parameters of the convective PBL simulation are reported in 
Table I. The Lagrangian analysis has been performed on an ensemble of numerical 
particle pairs, deployed when the LES has reached the quasi-steady state, 
 after six turnover times from the initialization (see Gioia et al., 2004, for more details about 
 this convective simulation).

\subsection{The LES Lagrangian experiments}
\label{sec:lagdisp}
In what might be considered the first application of LES to 
particle dispersion, Deardorff and Peskin (1970) reported the
Lagrangian statistics of one and two particle displacements for a 
LES turbulent channel flow. In spite of the relatively low resolution
and low number of particles the computed mean square particle
displacements were found to be consistent with Taylor's (1921)
theory. A more detailed investigation of Lagrangian particle
dispersion in a  convective boundary layer (CBL) has been conducted
later by Lamb (1978, 1979, 1982). His formulation for Lagrangian
diffusion model to calculate ensemble mean concentration involves a 
probability density function (pdf) of particle displacements. 
He obtained this pdf from a large ensemble of trajectories. 
The single particle trajectory is given by:
\begin{equation}
	\frac{dr_i^{(n)} }
{dt} = \bar V_i \left[ {\vec{r}^{(n)} \left( t \right),t} \right] + 
V'_i \left[ {\vec{r}^{(n)} \left( t \right),t} \right]
\label{eq:single_traj}
\end{equation}
where $r_i^{(n)}$ denotes the $i-$th component of the position vector $\vec{r}^{(n)}$ 
of the $n$-th particle, 
$\bar V_i$ represents the $i$-th component of the resolved wind field given by the LES and 
$V'_i$ represents the $i$-th component of the subgrid velocity component. 
The procedure by which  
$V'_i$ is determined is described in detail by Lamb (1981, 1982). 
From this model Lamb calculated the ensemble mean concentration and
reproduced the well know convection tank experiments by 
Willis and Deardorff (1976, 1981).

The energy flux within the inertial range of the LES, under well developed turbulence conditions, 
is computed as (Moeng, 1984, Sullivan et al., 1994):
\begin{equation} 
 \epsilon(x,y,z)  = \left( {0.19 + 0.74\frac{l}
{{\Delta s}}} \right)\frac{\bar e(x,y,z)^{3/2}}{l}
\end{equation}
where $\bar e(x,y,z)$ is the subgrid scale energy, 
$\Delta s = (\Delta x \Delta y \Delta z)^{1/3}$,
$\Delta x$,  $\Delta y$ and $\Delta z$ being the grid-spacing along
the three axes, and $l=\Delta s$.  
The mean value $\epsilon$ is obtained as
\begin{equation}
 \epsilon  = \frac{1}
{{0.3z_i L_y L_x}}\int\limits_{0.2z_i }^{0.5z_i} \int\limits_{0}^{L_y} \int\limits_{0}^{L_x} 
{\epsilon(x,y,z) dxdydz} 
\end{equation}
where $L_x$ and $L_y$ are the horizontal edges of the domain and $z_i$ is the mixing layer height.\\ 
Notice that, in the computation of the mean energy flux, we discard the highest
values of $\epsilon(x,y,z)$ close to ground for the
well-known limitations of the LES strategy in the vicinity of the  
wall boundaries. The mean energy flux within the LES inertial range, $\epsilon=\epsilon_{LES}$,  
defines the equivalent mean energy flux for the DSF inertial range, $\epsilon=\epsilon_{DSF}$,  
in the LES+DSF coupled model, as discussed in the next section. Another important parameter to use for 
the LES+DSF coupling is the LES (horizontal) 
grid step, indicated as $\Delta L=\Delta x=\Delta y$. 
 
Once the quasi-steady regime is settled, after the initial transient phase,  
we have seeded the LES flow  
with 4096 particle pairs, uniformly distributed on a horizontal plane, advected in time (in
parallel with the LES model and with the same time-step $dt=1$ $s$) according to 
Eq.~(\ref{eq:single_traj}) for 4500 time-steps.   
As in Gioia et al. (2004), 
the knowledge of the velocity field in any point, necessary to integrate (\ref{eq:single_traj}),
is obtained by a bilinear interpolation
of the eight nearest grid points in which the winds generated 
by the LES are explicitly defined. 
The details of the LES Lagrangian experiments (with or without the subgrid kinematic model) 
are reported in Tab. I.

\section{Results and discussion}
\label{sec:results}

\begin{figure}[h]
	\begin{center}
		\includegraphics[angle=0,width=15cm,height=15cm]{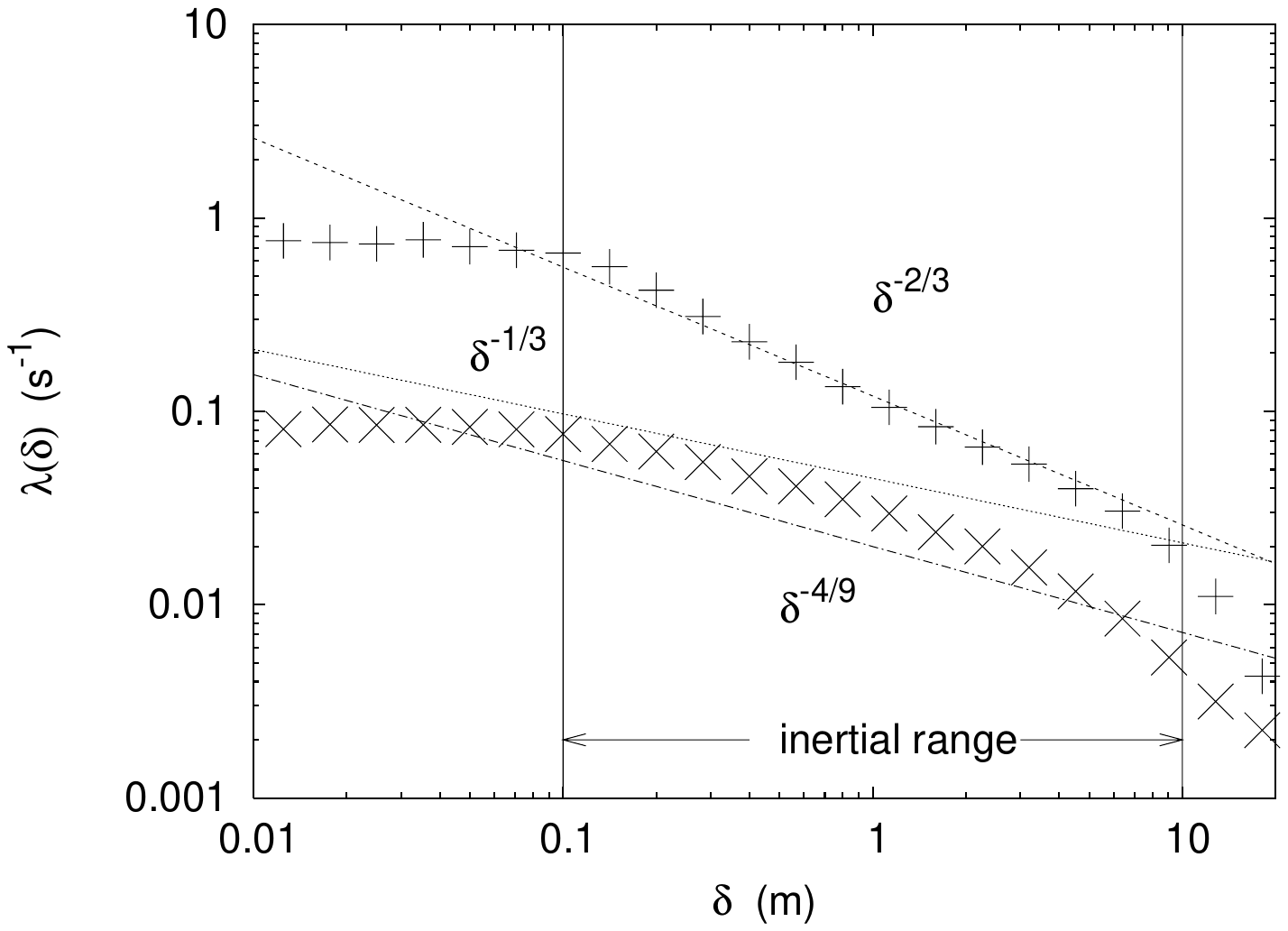}
	\end{center}
		\caption{}
\label{fig:fsle_km}
\end{figure}
\begin{figure}[h]
	\begin{center}
		\includegraphics[angle=0,width=15cm,height=15cm]{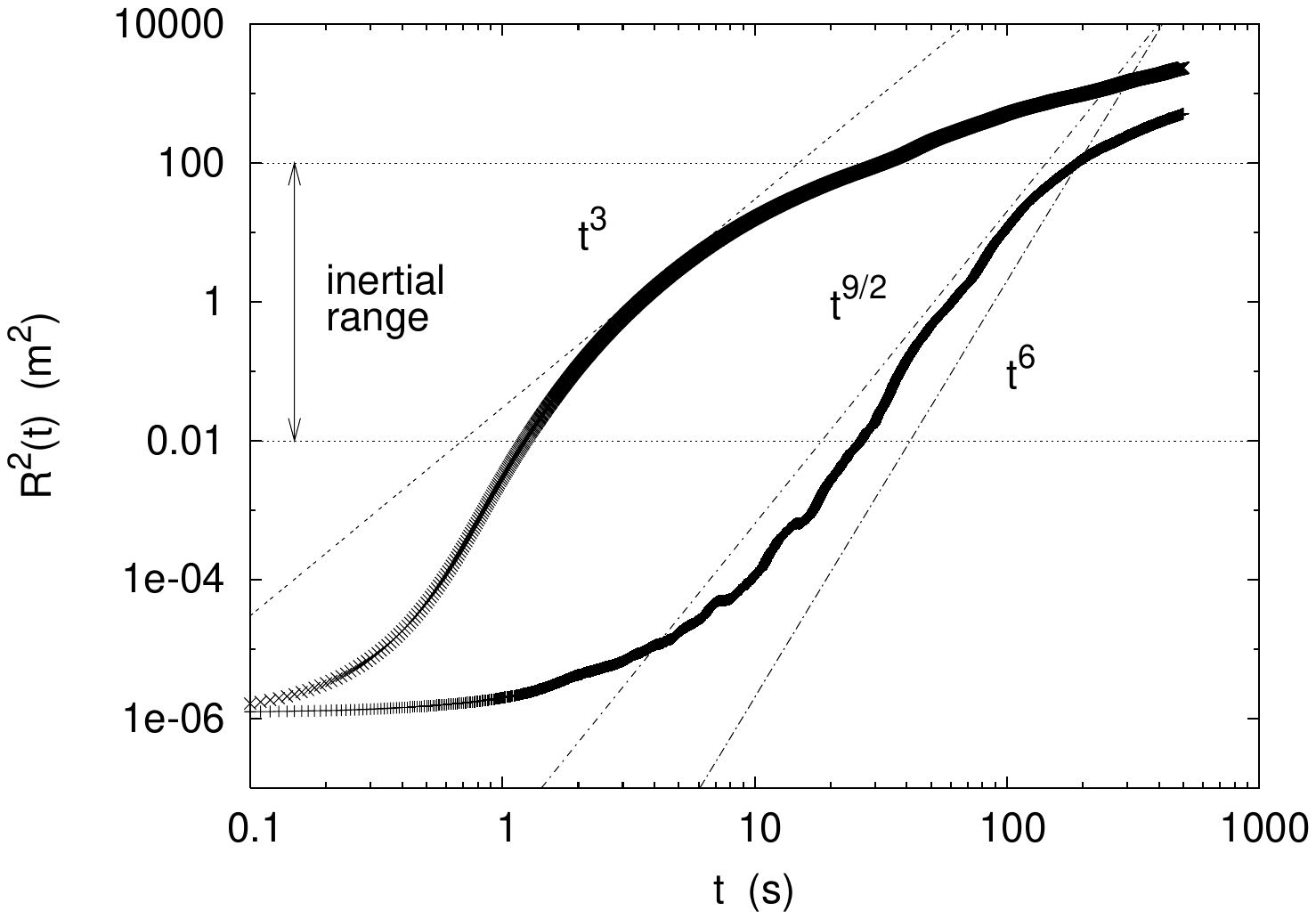}
	\end{center}
		\caption{}
\label{fig:rd_km}
\end{figure}

Let us consider, first, the behavior of the DSF model, as far as relative dispersion is concerned, 
and see how the relative coordinates correction allows to reproduce the expected Richardson's scaling 
within the inertial range. A uniform drift 
field, $V_d \gg A^{(1)}$, is added to simulate the presence of a strong mean advection, 
and test the response of the model against 
the Thomson and Devenish (2005) predictions for the no "sweeping effect" case. 
Given the parameter set-up reported in Tab. II, the inertial range 
 approximately corresponds to the interval $[l_1^{(N_m)}/2,l_1^{(1)}/4]=[0.1,10]$ $m$, 
 for the geometrical properties of the flow discussed before. As can be numerically shown,   
 the way the upper (lower) bound of the inertial range are related to the largest (smallest) wavelength, 
 respectively, depends neither on the width of the inertial range nor on the density of the modes. 
 The equivalent mean energy flux, $\epsilon_{DSF}$, is set to $10^{-3}$ $m^2s^{-3}$.  
 It can be also verified that, of course, the behavior of the model does not depend  
 on the total energy of the modes. 
 The constant $C_K$ is set to $0.25$, such that the corresponding 
 Richardson's constant has a value $C_R \simeq 0.5$, computed from the FSLE $\delta^{-2/3}$ scaling 
 (Boffetta and Sokolov, 2002). It can be shown numerically that $C_R$ does not change  
 if $\epsilon_{DSF}$ is varied over many orders of magnitude.  
 The choice of this particular value for $C_R$ is rather conventional, we 
 consider, indeed, values of the order $10^{-1}$, i.e. within the $[0.1,1]$ interval,  
 as acceptable, according to the most recent estimates of the Richardson's constant.    
 The integration time step, $dt$, is chosen sufficiently 
 small in order to test the response of the DSF model to the "sweeping effect". In particular, if 
 we indicate with $A_{rms}=\langle A^2 \rangle^{1/2}$ the root mean square velocity averaged 
 over all the $N_m$ modes, and with $A_{max}=A^{(1)}+A^{(2)}+...+A^{(N_m)}$ the maximum available velocity 
 in one point, a value $dt=10^{-2}$ $s$ turns out to be smaller than all the shortest characteristic advective 
 times across the smallest eddies (i.e. eddies of size $\simeq 0.1$ $m$), 
 even considering the additional drift velocity, see Table II.  
 In Figs. \ref{fig:fsle_km} and \ref{fig:rd_km}, the results concerning the FSLE, function of the particle separation,  
 and the mean square relative dispersion, 
 function of the time interval from the release, are reported. Both cases, with "sweeping effect" correction and 
 without "sweeping effect" correction, have been analysed. It is shown quite clearly, 
 especially by means of the FSLE, how the "relative coordinate" technique 
 allows to recover the right relative dispersion behavior, in agreement with the $\delta^{-2/3}$ Richardson's law, 
  against the anomalous scalings, $\delta^{-4/9}$ and $\delta^{-1/3}$, appearing if no "sweeping effect" is 
  taken into account, as predicted by Thomson and Devenish (2005). The same picture holds in terms of the mean square 
  relative displacement, being in this case $t^3$, $t^{9/2}$ and $t^6$ the 
  equivalent scaling laws to match with the data, as can be verified by dimensional arguments. 
  We would like to precise that 
  we are not concerned, in this context, in verifying which of the Thomson and Devenish (2005) predictions best 
  fits the results obtained with no "sweeping effect" correction. We are mainly interested, instead, 
  in the appearance of the Richardson's scaling when the "sweeping effect" correction is taken into account, 
  regardless of the intensity of the mean drift velocity added to the DSF field.   
  It is also to be remarked, once again, that the finite-scale dispersion rate statistics, 
  based on the FSLE, provides more physically consistent 
  informations than what results, generally, from the statistics based on the time growth of the mean square relative displacement, for the reasons explained above. 
  
\begin{figure}[h]
	\begin{center}
		\includegraphics[angle=0,width=15cm,height=15cm]{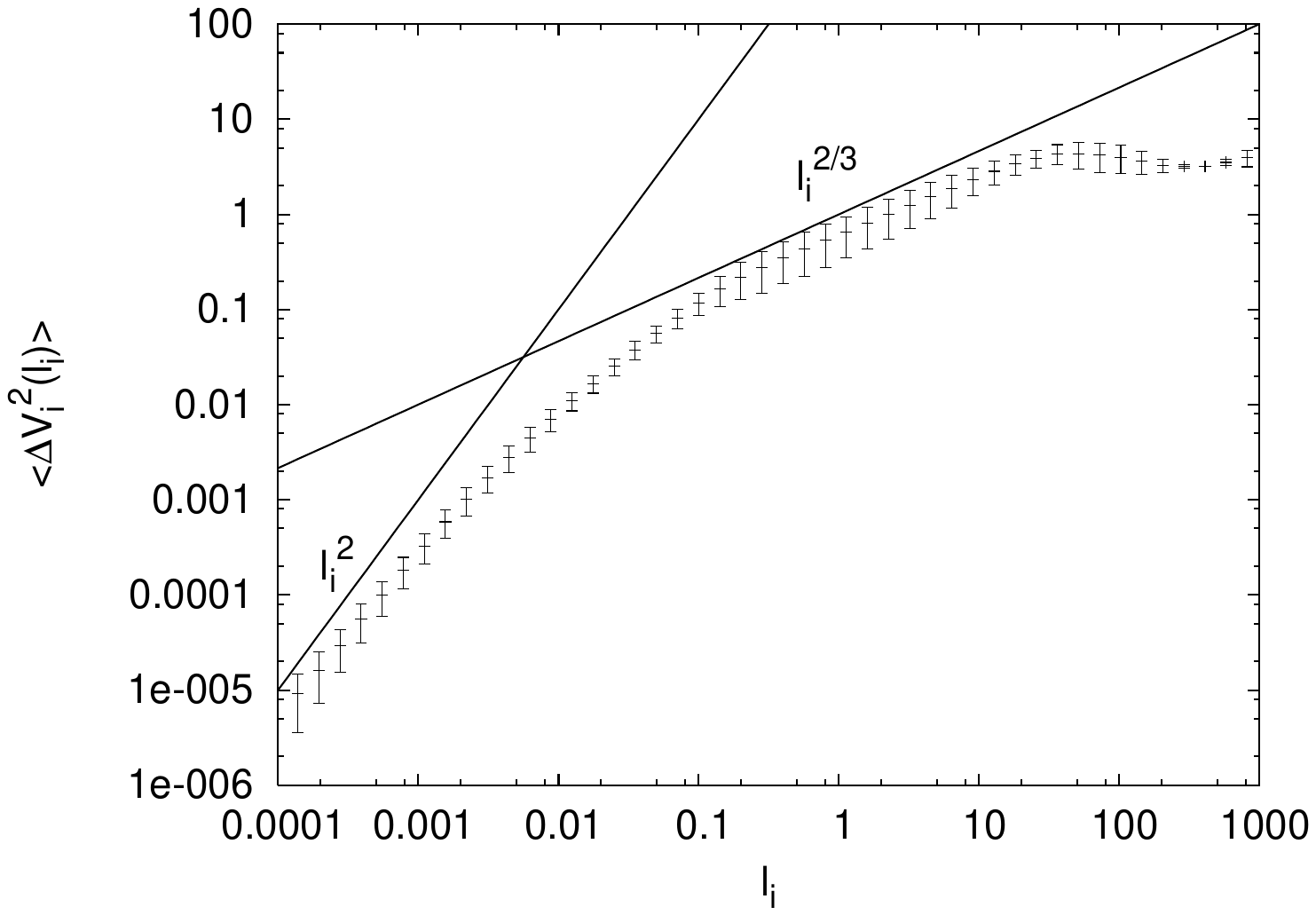}
	\end{center}
	\caption{}
\label{fig:lsf_km}
\end{figure}

\begin{figure}[h]
	\begin{center}
		\includegraphics[angle=0,width=15cm,height=15cm]{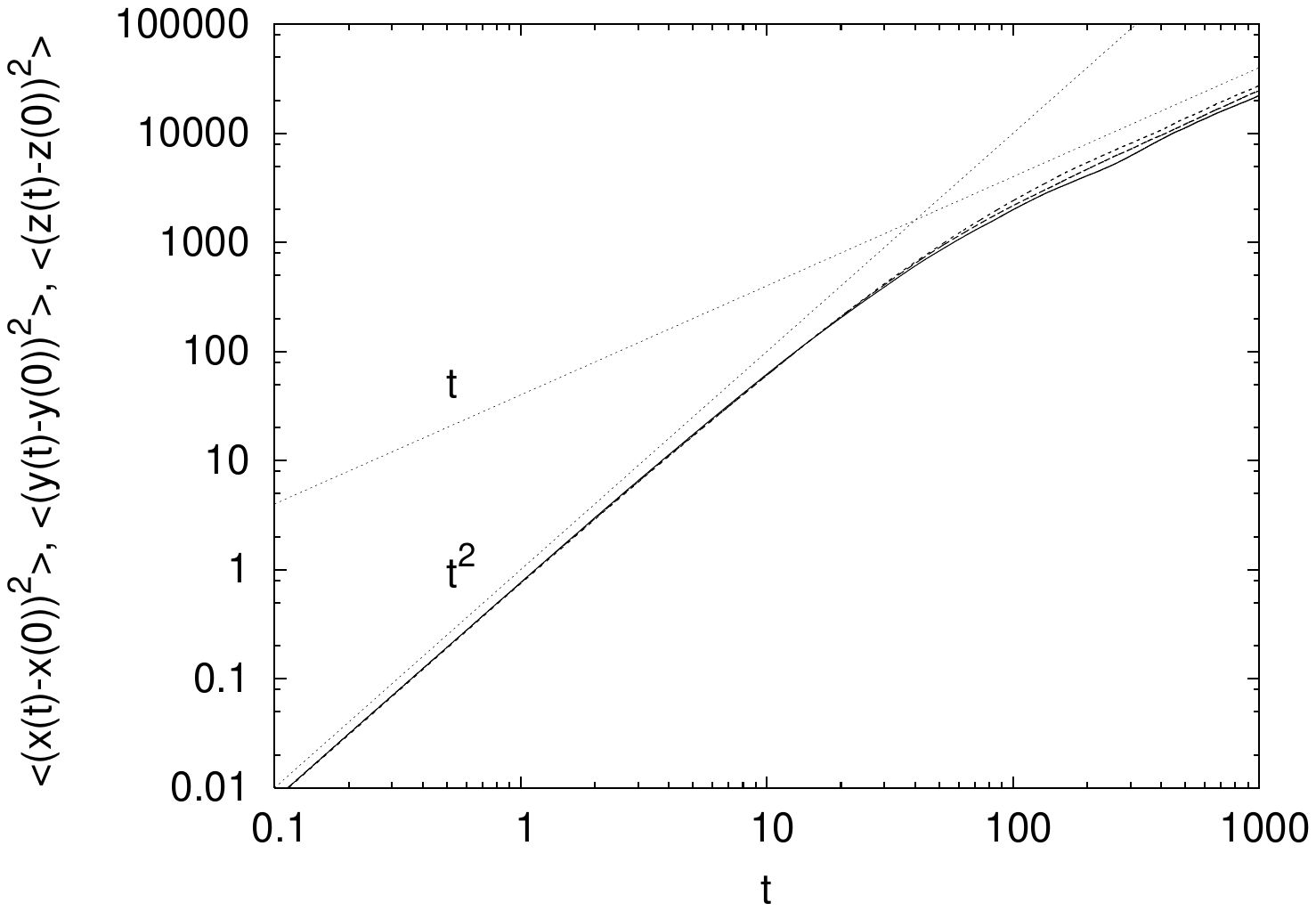}
	\end{center}
	\caption{}
\label{fig:ad_km}
\end{figure}

\begin{figure}[h]
	\begin{center}
		\includegraphics[angle=0,width=15cm,height=15cm]{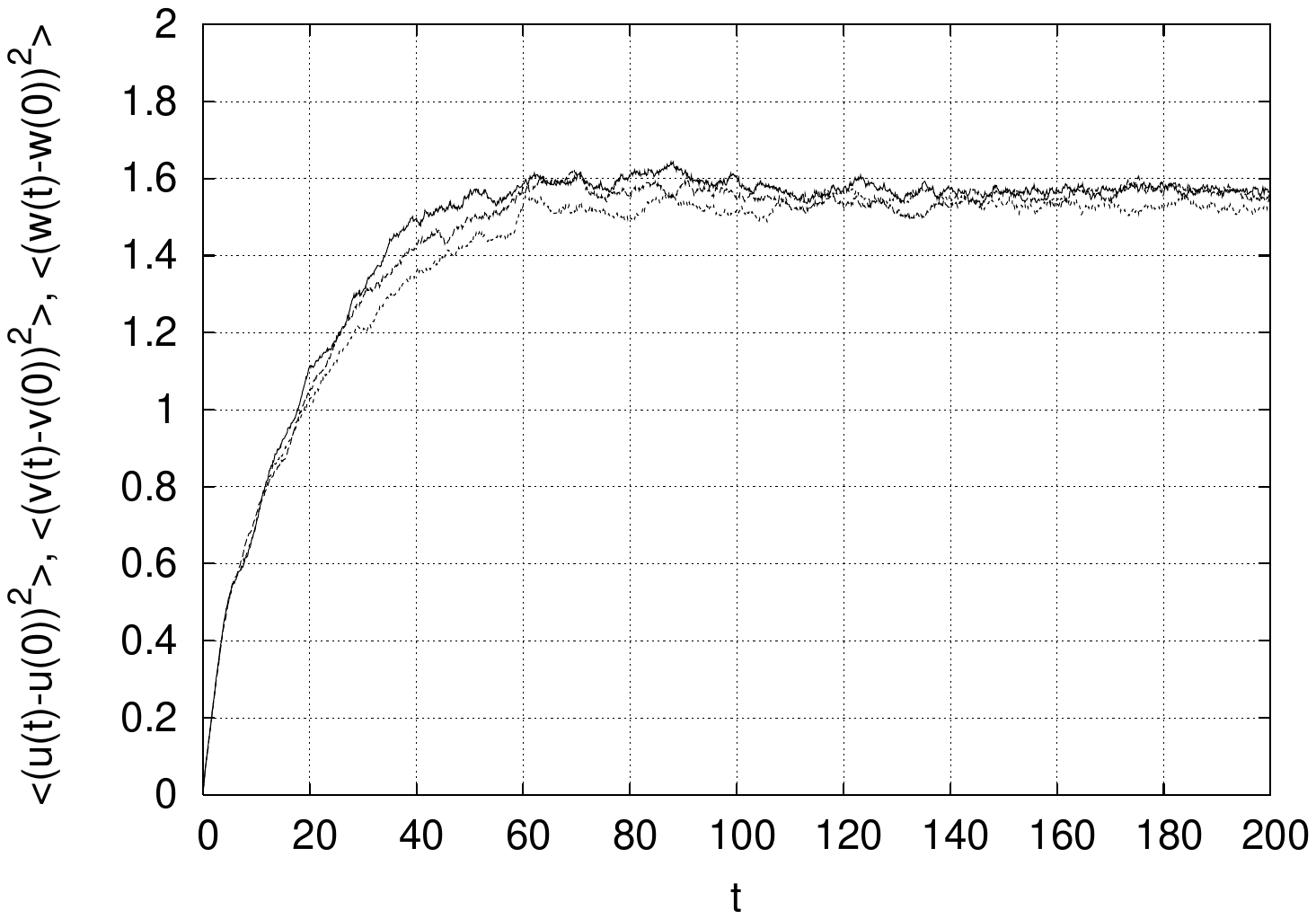}
	\end{center}
	\caption{}
\label{fig:sf2_km}
\end{figure}

\begin{figure}[h]
	\begin{center}
		\includegraphics[angle=0,width=7.cm,height=7.cm,angle=-90.]{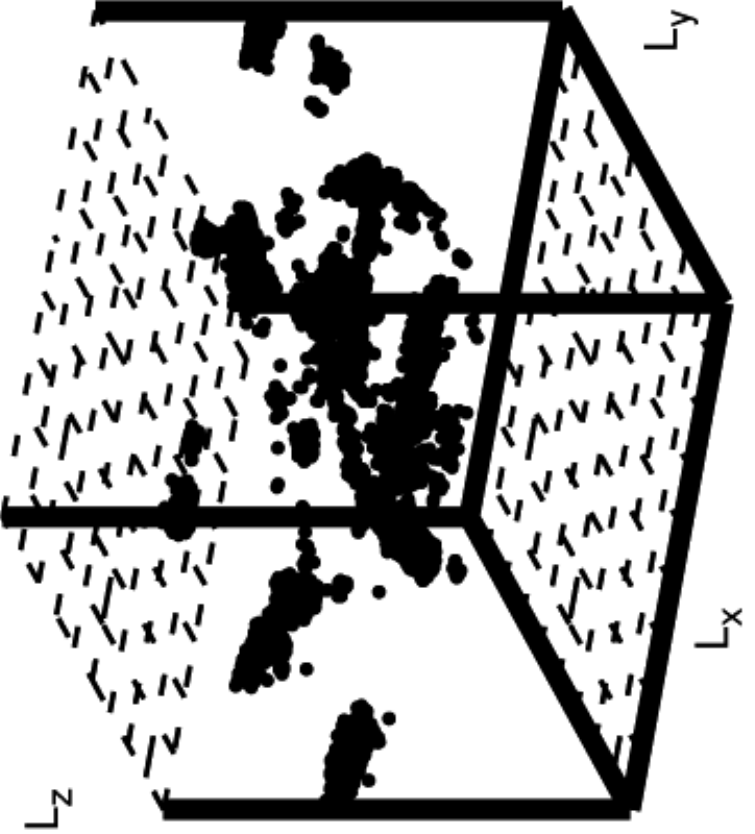}
		\includegraphics[angle=0,width=7.cm,height=7.cm,angle=-90.]{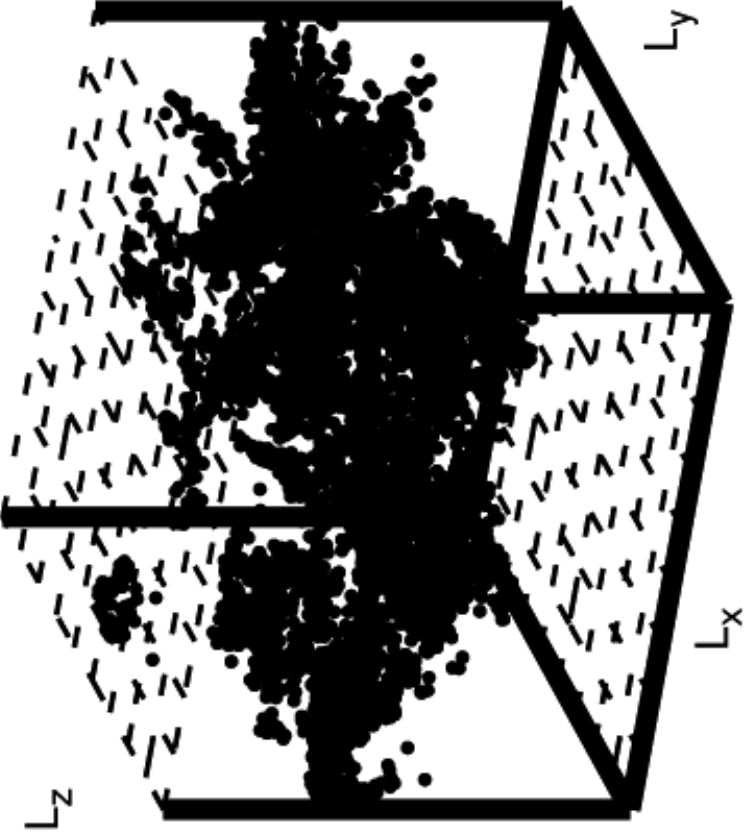}
		\includegraphics[angle=0,width=7.cm,height=7.cm,angle=-90.]{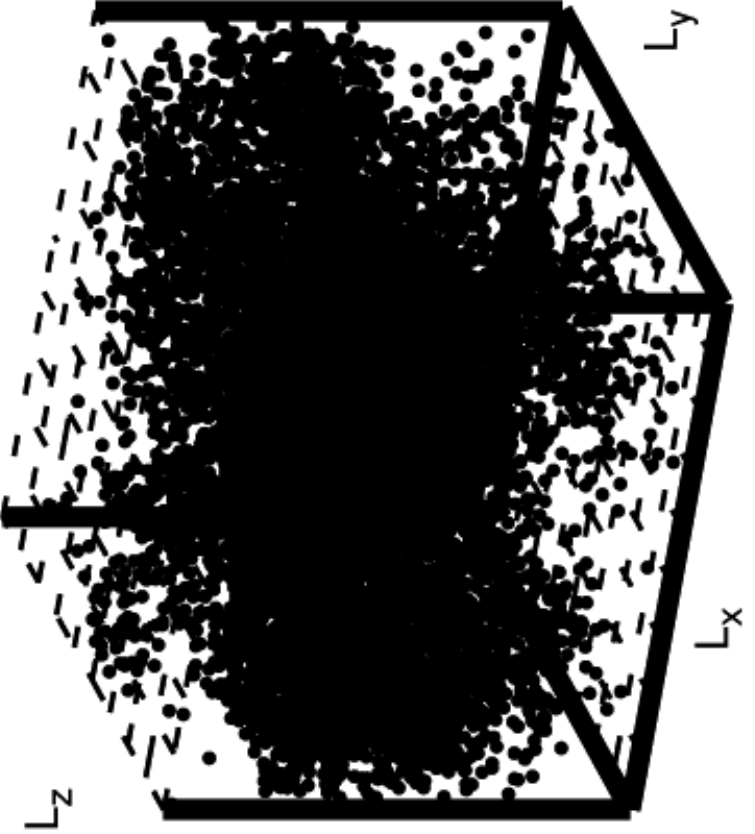}
		\includegraphics[angle=0,width=7.cm,height=7.cm,angle=-90.]{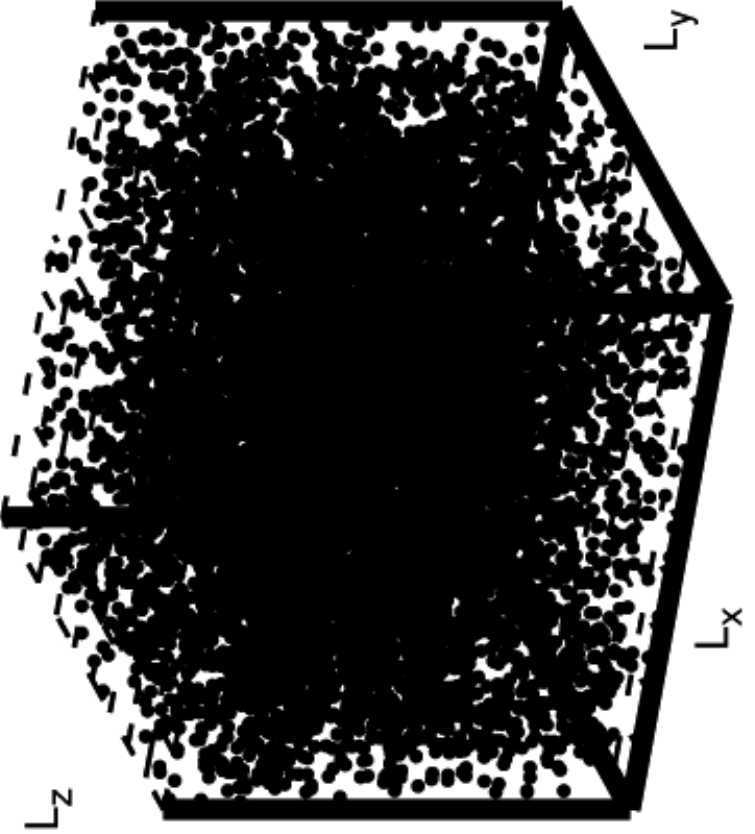}
	\end{center}
	\caption{}
\label{fig:mixing_km}
\end{figure}

Once the improvement, provided by the relative coordinates technique, to turbulent dispersion kinematic modeling 
has been established, we will use, henceforth, the DSF model as defined 
by (III.13), (III.14) and (III.15), i.e. with the "sweeping effect" correction incorporated. 
We have performed further analysis of the DSF model, with a set-up described in Tab. III, in order to examine 
the isotropy properties of Lagrangian dispersion. Let us consider, first, another two-particle statistics, 
as support to the FSLE computation. If we indicate with $l_i$ (for $i=1,2,3$)  
a component of the separation $\Delta \vec{r}$ between two particles, at a fixed time,  
along a certain direction, and with $\Delta V_i$ the correspondent component of the 
Lagrangian velocity difference, along the same direction, then, according to Kolmogorov's theory (Frisch, 1995),  
 we expect $\Delta V_i(l_i)^2 \sim l_i^{2/3}$. 
 In Fig. \ref{fig:lsf_km} we report the mean square velocity 
difference, averaged over the three directions, as function of the correspondent component of the 
distance between two particles. The Kolmogorov scaling, inside the inertial range of the 
DSF model ([$0.1,10^2$] $m$), is an indirect confirmation of the existence of the Richardson's law as 
regards to relative dispersion. 
As far as the one-particle statistical properties of the flow are concerned, we have computed the 
three components of the mean square absolute dispersion of an ensemble of particles, 
reported in Fig. \ref{fig:ad_km}, 
the three components of the second-order temporal structure function of the 
Lagrangian velocity along a trajectory, reported in Fig. \ref{fig:sf2_km}, and 
the evolution of the spatial distribution of an ensemble of particles, sampled at four different fractions 
of the largest turnover time of the flow, reported in Fig. \ref{fig:mixing_km}.  
The one-particle statistics show a low anisotropy degree, not higher than $10-20\%$. 
These results justify the fact of setting the ratio between vertical and horizontal wavenumber, for 
each mode, equal to $2$. 
Absolute dispersion show, as expected, the existence of the two regimes $\sim t^2$ and $\sim t$, for, 
respectively, short times and long times, typical of diffusion motion with finite-time autocorrelations. 
The mean square time-delayed velocity difference, along the single trajectories, is characterized 
by an initially isotropic linear growth which later slowly approach a saturation level, 
on a time scale comparable 
to the turnover times of the largest modes, where small differences among the three directions due 
to anisotropy effects become more visible.       
The mixing properties of the flow are evident when looking at the approximately uniform dispersion  
of a cloud of particles, initially distributed along a horizontal line, at half height of a box 
of edges $L_x=L_y=L_z=l_1$, on a time scale of the order of the turnover time. The three spatial coordinates 
of the particles are imposed to have periodic boundary conditions with respect to the box.    
      
\begin{figure}[h]
	\begin{center}
		\includegraphics[angle=0,width=15cm,height=15cm]{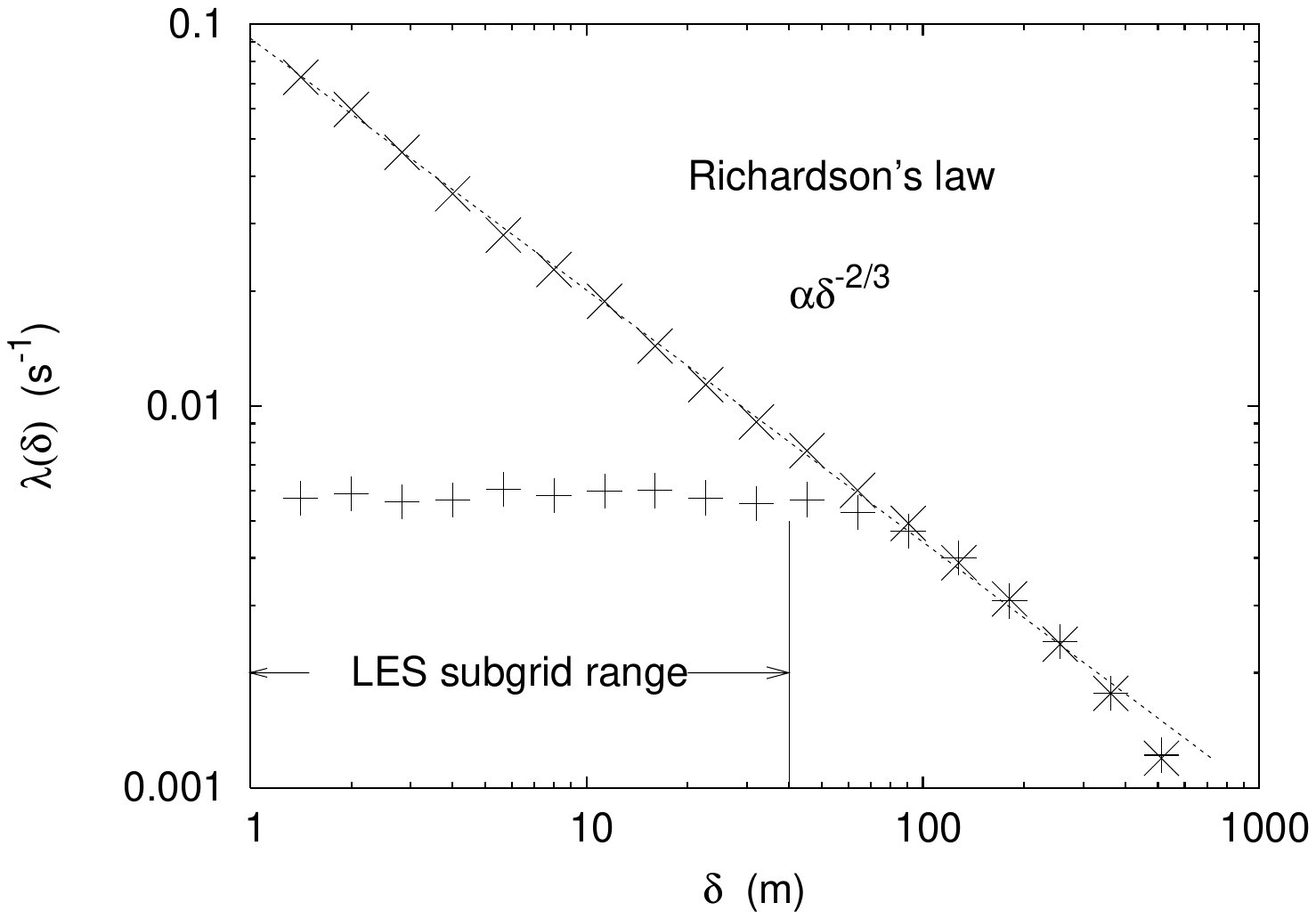}
	\end{center}
		\caption{}
\label{fig:fsle_lesdsf}
\end{figure}

\begin{figure}[h]
	\begin{center}
		\includegraphics[angle=0,width=15cm,height=15cm]{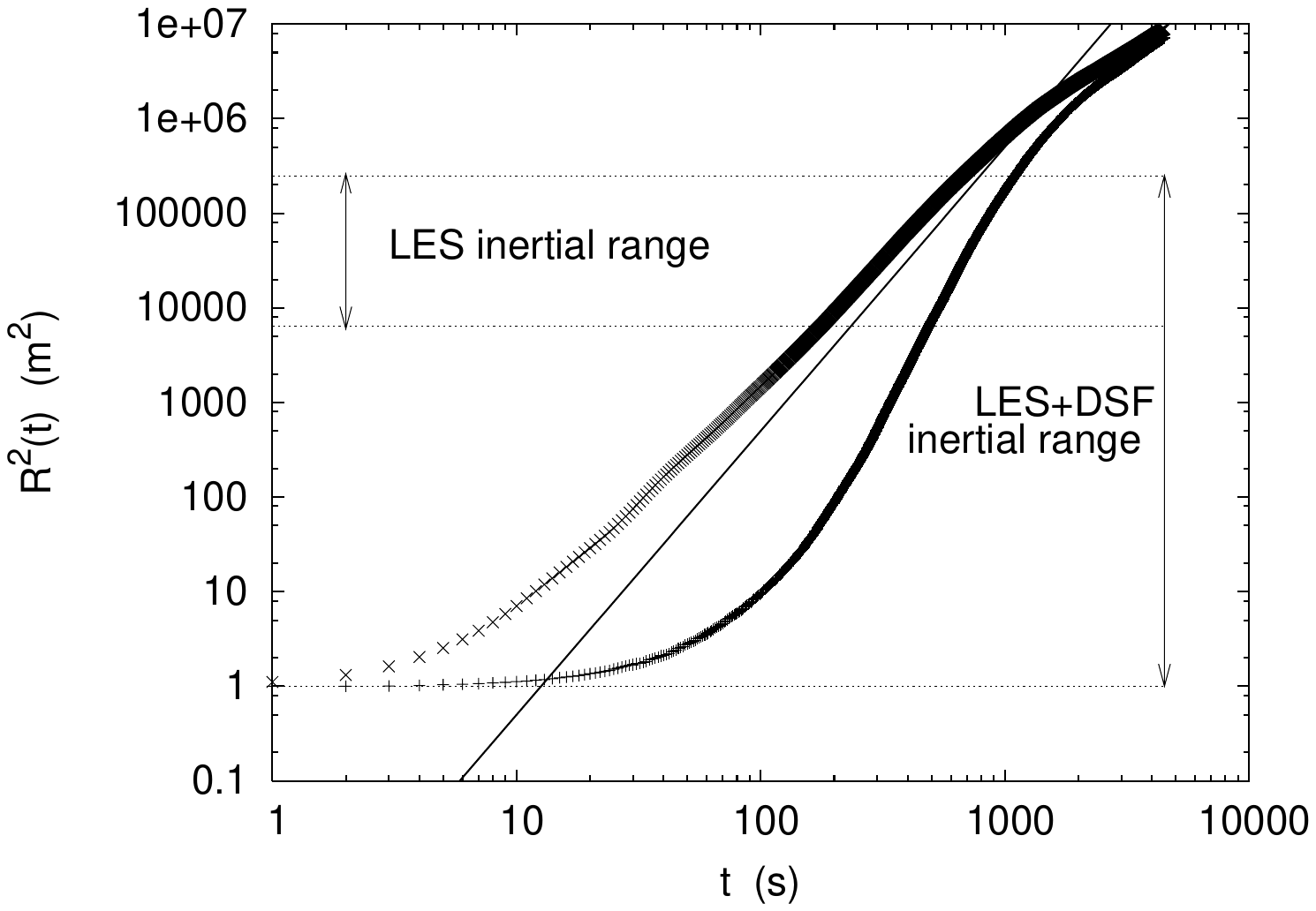}
	\end{center}
		\caption{}
\label{fig:rd_lesdsf}
\end{figure}

\begin{figure}[h]
	\begin{center}
		\includegraphics[angle=0,width=15cm,height=15cm]{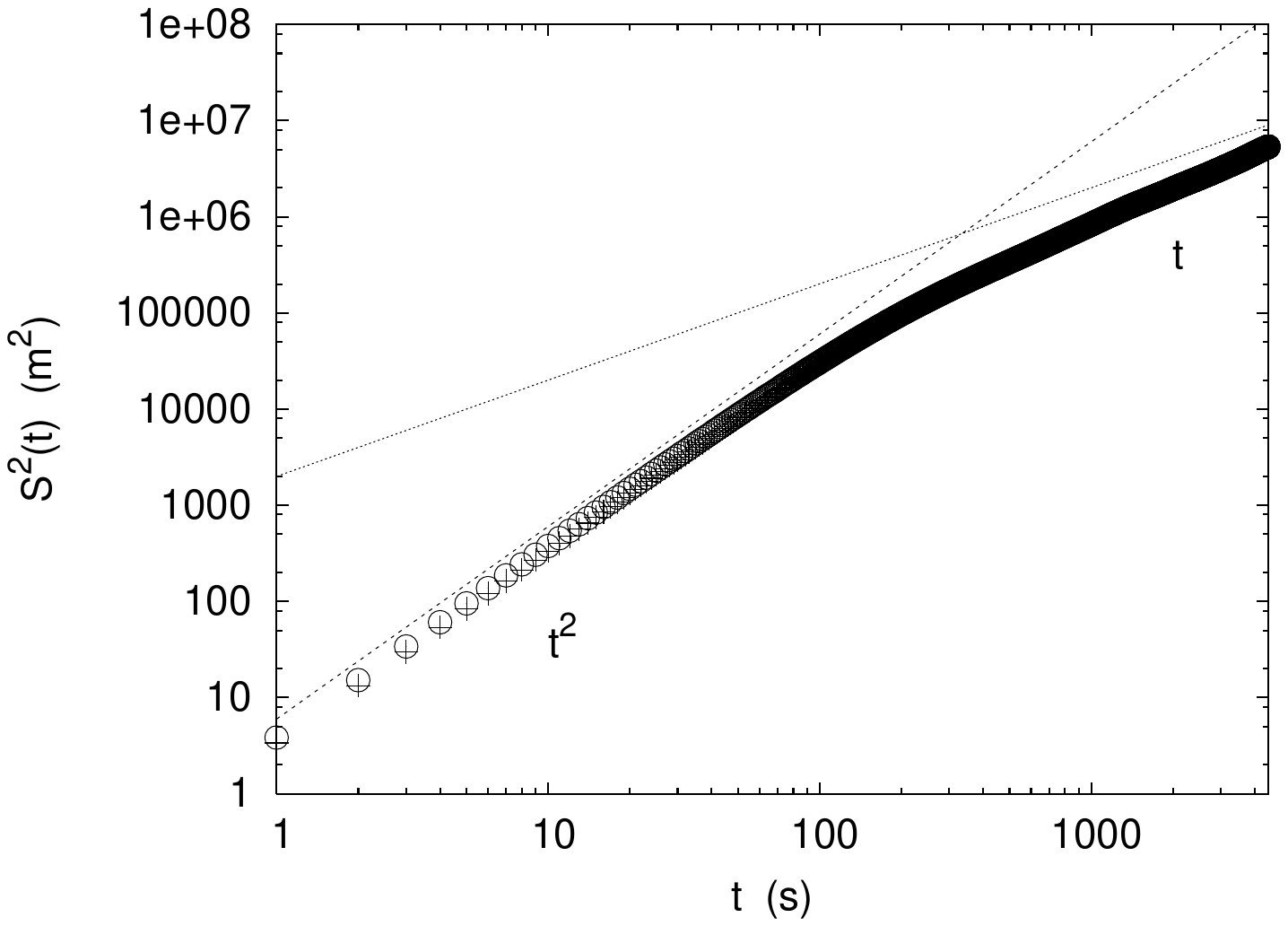}
	\end{center}
		\caption{}
\label{fig:ad_lesdsf}
\end{figure}

We will discuss now the results about the application of this model    
as subgrid kinematic field in the convective LES Lagrangian experiment (see Tab. I). 
In Fig. \ref{fig:fsle_lesdsf} the FSLE statistics, 
computed for the LES trajectories with and without subgrid coupling, is reported. 
An ensemble of $4096$ particle pairs is released, uniformly distributed, on a horizontal plane 
at about middle height of the domain, with initial particle separation $\delta_0=1$ $m$. 
The total simulation time of the trajectories is about nine times the LES turnover time. 
In absence of coupling, 
the convective LES is characterized by an inertial range starting approximately at $2\Delta L$, where 
$\Delta L$ is the LES (horizontal) grid step, and ending at a scale of about $4-500$ $m$, of the order of the 
mixing layer height of the PBL in quasi-steady convective regime. Subgrid relative dispersion, at  
$\delta < 2\Delta L$, is exponential with a constant growth rate given by the plateau level 
$\lambda \simeq 5 \cdot 10^{-3}$ $s^{-1}$. Since the subgrid velocity components of the LES are filtered out, 
this value of the LLE, $\lambda$, underestimate the actual dispersion rates as long as the particle separation 
remains smaller than the grid step scale. The coupling with the DSF model, used as subgrid kinematic field, 
allows to improve the description of the dispersion process, and to extend, in some sense, the LES inertial range 
from upgrid scales to subgrid scales, as shown in Fig. \ref{fig:fsle_lesdsf}. 
The set-up of the subgrid DSF model, see Tab. IV, is such that: the mean energy flux is the same for both models, 
$\epsilon_{DSF}=\epsilon_{LES}$; the first mode wavelength is $l_1^{(1)}=8\Delta L$ so that the upper bound of 
the kinematic inertial range corresponds to $2\Delta L$; the last mode wavelength is fixed by the condition 
that the smallest eddy turnover time, $\tau^{(N_m)}$, must be at least one order of magnitude larger than 
the integration time step ($dt=1$ $s$); the number of kinematic modes $N_m$ is then 
fixed by the parameter $\rho=2^{1/4}$; the constant $C_K$ is suitably tuned to the value $0.125$ so to have 
a smooth transition of the FSLE from upgrid scales to subgrid scales. We verified that integrating the DSF field 
at a time step shorter than the LES time step does not carry substantial improvements, while    
the only, unwanted, effect is to increase the computational time. 
The Richardson's law $\alpha \delta^{-2/3}$ is rather well compatible with the data, and the value of the 
Richardson's constant, estimated from $\alpha$ (Boffetta and Sokolov, 2002), is  
$C_R \sim 10^{-1}$. The action of the subgrid kinematic model, now, allows to observe dispersion rates even 
an order of magnitude larger than the LLE found for the uncoupled LES. 
In Fig. \ref{fig:rd_lesdsf}, the same picture, in terms of the mean square particle displacement $R^2(t)$, is 
reported. The $t^3$ Richardson's law is plotted against the two curves to show that, for the coupled model 
LES+DSF, relative dispersion has a major tendency to follow the expected behavior than for the uncoupled LES. 
The "memory effect" of the initial conditions is visible through the presence of a transient time before the 
Richardson's scaling is approached. The standard diffusion regime begins after a time interval 
 about four times longer than the LES turnover time. Using the same set of $4096$ particle pairs, it is possible to 
 measure absolute dispersion too, averaged over all the $2 \, \times \, 4096$ single trajectories. 
 In Fig. \ref{fig:ad_lesdsf}, the behavior of the 
 two-time, mean square dispersion from the initial release point is reported.  
 It can be verified that the presence of the subgrid DSF field does not affect 
 the absolute dispersion properties of the LES, as expected. The asymptotic standard diffusive regime 
 begins after a time interval about four times the LES turnover time.

\section{Conclusions}
\label{sec:conclusions}

A $3D$ non linear, deterministic velocity field, 
derived from a double stream-function, named DSF model, has been introduced and discussed as a 
kinematic simulation for modeling Lagrangian turbulent particle dispersion. 
Multiscale chaotic dynamics is the mechanism that generates turbulent-like trajectories from a non turbulent velocity field. The DSF model is made of eddies ($3D$ non steady convective cells) that are kept fixed in space, 
exception made for 
the periodic oscillations around their equilibrium positions, and with infinite lifetime, i.e. non decaying 
one-point Eulerian correlations. 
Eulerian turbulence, of course, cannot be modeled in realistic terms by means of kinematic simulations, but we have 
shown that, if the velocity amplitude of the modes 
is related to the spatial wavelength through the Kolmogorov scaling, and if the lack of "sweeping effect" of the 
small eddies by the large eddies, typical of kinematic simulation (Thomson and Devenish, 2005), 
is overtaken, the DSF model can reproduce correctly the expected Lagrangian 
dispersion properties of a turbulent flow. 
If we consider the DSF model as a two-particle dispersion model, 
the "sweeping effect" can be simulated by replacing the particle absolute coordinates with  
the relative coordinates to their mass center. This is done for all modes except the first one, so that there 
still exists a large scale mixing that allows the spatial average of the dispersion process. This technique 
assures that eddies of every size (except the largest one) move anchored to the mass center of a particle pair, 
while, at the same time, make the two particle separate form each other according to the Richardson's law. 
There is no relative shift among the eddies, in the sense that all the eddies (except the largest ones) 
are advected at the same speed, together with the mass center of a particle pair, but, what is most important, 
is that the "fast crossing" of the particle pairs through the eddies, caused by the large scale advection, is in this 
way eliminated. 
On the basis of its properties, the DSF model can be successfully used as subgrid kinematic field in LES 
Lagrangian experiments, provided that the mean energy flux is the same in the two models, the upper 
bound of the DSF inertial range correspond to the lower bound of the LES inertial range and the $C_K$ 
parameter in the DSF model is suitably tuned to grant a smooth transition from upgrid to subgrid scales. 
All the procedure is consistent with an estimate of the Richardson's constant of the order 
$C_R \sim 10^{-1}$, in agreement with the most recent results on both experimental and numerical 
turbulent flows.

\bigskip

\begin{acknowledgments} 
We would like to warmly thank two anonymous Referees for their constructive criticism, 
G. Gioia for her contribution to the early stage of this work, A. Moscatello for her help with the graphics 
and A. Vulpiani for helpful discussions and suggestions about the DSF kinematic model.  
This work has been supported by COFIN 2005 project
n.~2005027808 and by CINFAI consortium (A.M.).

\end{acknowledgments}

\newpage

\parindent -0.5in

\centerline{REFERENCES}

Antonelli, M., A. Mazzino, and  U. Rizza, 2003:
Statistics of temperature fluctuations
in a buoyancy dominated boundary layer flow simulated by a
Large-eddy simulation model.
{\it J. Atmos. Sci.}, {\bf 60}, 215--224.

Aurell, E., G. Boffetta, A. Crisanti, G. Paladin and A. Vulpiani, 1996:
Growth of non-infinitesimal perturbations in turbulence. 
{\it Phys. Rev. Lett.}, {\bf 77}, 1262--1265.

------ , G. Boffetta, A. Crisanti, G. Paladin, and A. Vulpiani, 1997: 
Predictability in the large: an extension of the concept of Lyapunov 
exponent. 
{\it J. Phys. A: Math. Gen.}, {\bf 30}, 1--26.

Artale, V., G. Boffetta, A. Celani, M. Cencini, and A. Vulpiani, 1997:
Dispersion of passive tracers in closed basins: beyond the diffusion 
coefficient. 
{\it Phys. of Fluids}, {\bf 9}, 3162--3171.

Boffetta, G., A. Celani, M. Cencini, G. Lacorata, and A. Vulpiani, 2000:
Non-asymptotic properties of transport and mixing.   
{\it Chaos}, {\bf 10}, 1, 1--9.

Chirikov, B. V., 1979: 
A universal instability of many dimensional oscillator systems. 
{\it Phys. Rep.}, {\bf 52}, 263--379.

Crisanti, A., M. Falcioni, G. Paladin, and A. Vulpiani, 1991: 
Lagrangian Chaos: Transport, Mixing and Diffusion in Fluids. 
{\it Nuovo Cimento}, {\bf 14}, n. 12, 1--80.

Deardorff, J. W., 1972:
Numerical investigation of the planetary boundary layer with an inversion lid.
{\it J. Atmos. Sci.}, {\bf 29}, 91--115. 

------ , and Peskin, R. L., 1987: 
Lagrangian statistics from numerical integrated turbulent shear flow. 
{\it Phys. of Fluids}, {\bf 13}, 584--595.

Frisch, U., 1995: 
Turbulence: the legacy of A.N. Kolmogorov. 
{\it Cambridge Univ. Press}, pp. 310.

Fung, J. C. H., J. C. R. Hunt, N. A. Malik and R. J. Perkins, 1992: 
Kinematic simulation of homogeneous turbulence by unsteady random Fourier modes. 
{\it J. Fluid Mech.}, {\bf 236}, 281--318.

------ , and J. C. Vassilicos, 1998: 
Two particle dispersion in turbulent-like flows. 
{\it Phys. Rev. E}, {\bf 57}, n. 2, 1677--1690.

Gioia, G., G. Lacorata, E. P. Marques Filho, A. Mazzino, and U. Rizza, 2004: 
The Richardson's Law in Large-Eddy Simulations of Boundary Layer flows. 
{\it Boundary Layer Meteor.}, {\bf 113}, 187--199.

Joseph, B., and B. Legras, 2002:
Relation between kinematic boundaries, stirring and barriers
for the antarctic polar vortex.
{\it J. of Atmos. Sci.}, {\bf 59}, 1198--1212.

LaCasce, J. H., and C. Ohlmann, 2003: 
Relative dispersion at the surface of the Gulf of Mexico. 
{\it J. Mar. Res.}, {\bf 61}, 285--312.

Lacorata, G., E. Aurell, and A. Vulpiani, 2001:
Drifter dispersion in the Adriatic Sea: Lagrangian data and chaotic model.
{\it Ann. Geophys.}, {\bf 19}, 121--129.

------ , E. Aurell, B. Legras, and A. Vulpiani, 2004:
Evidence for a $k^{-5/3}$ spectrum from the EOLE Lagrangian balloons in the low stratosphere.
{\it J. Atmos. Sci.}, {\bf 61}, 2936--2942.

Lamb, R. G., 1978: 
A numerical simulation of dispersion from an elevated point source in the convective boundary layer. 
{\it Atmos. Environ.}, {\bf 12}, 1297--1304. 

------ , 1979: 
The effects of release height on material dispersion in the convective planetary boundary layer. 
Preprint Vol., Fourth Symposium on turbulence, diffusion and air pollution. 
American Meteorological Society, Boston, 27--33.

------ , 1982: 
Diffusion in the convective boundary layer. In atmospheric turbulence and air pollution modeling. 
(F.T.M Nieuwstadt and H. van Dop, eds) D.Reidel Pub.Co., Dordrecth, Holland, 158--229.

Leonard, A., 1974: 
Energy cascade in large-eddy simulations of turbulent fluid flows. 
{\it Adv. in Geophysics, Academic Press}, {\bf 18}, 237--248.

Lichtenberg, A. J., and M. A. Lieberman, 1982: 
 Regular and stochastic motion.
{\it Springer-Verlag}, pp. 655.

Lilly, D. K., 1967: 
The representation of small-scale turbulence in numerical simulation experiments. 
{\it Proc. IBM Sci. Comput. Symp. Environ. Sci., IBM Data Process. Div., White Plains, N.Y.}, 195--210.

Lorenz, E., 1963:
Deterministic nonperiodic flow.
{\it J. of Atmos. Sci.}, {\bf 20}, 130--141.

Moeng, C.-H., 1984: 
A Large-Eddy Simulation model for the study of planetary boundary layer turbulence. 
{\it J. of Atmos. Sci.}, {\bf 41}, 2052--2062.

------ , and Wyngaard, J.C., 1988: 
Spectral analysis of Large-Eddy Simulations of the Convective Boundary Layer. 
{\it J. of Atmos. Sci.}, {\bf 45}, 3573--3587.

------ , and P.P. Sullivan, 1994: 
A comparison of shear and buoyancy driven planetary boundary layer flows. 
{\it J. of Atmos. Sci.}, {\bf 51}, 999--1021.

Ott, S., and J. Mann, 2000: 
 An experimental investigation of the relative diffusion of particle pairs in three-dimensional 
 turbulent flow. 
{\it J. Fluid Mech.}, {\bf 422}, 207--223.

Ottino, J. M., 1989: The kinematics of mixing: stretching, chaos and transport,
{\it Cambridge University Press }, pp. 378.

Porte Agel F., Meneveau C., Parlange M.B., 2000:
A scale-dependent dynamic model for large-eddy simulation: application to a neutral atmospheric boundary layer.
{\it J. Fluid Mech.}, {\bf 415}, 261--284.

Richardson, L. F., 1926:
Atmospheric diffusion shown on a distance-neighbor graph.
{\it Proc. R. Soc. London Ser. A}, {\bf 110}, 709--737.

Rizza, U., C. Mangia, J.C. Carvalho, and D. Anfossi, 2006:
Estimation of the Lagrangian velocity structure function constant C0 
by Large-Eddy Simulation.
{\it Boundary Layer Meteor.}, {\bf 120}, 25--37.

Sawford, B., 2001: 
Turbulent relative dispersion. 
{\it Ann. Rev. Fluid Mech.}, {\bf 33}, 289--317.

Solomon, T. H., and J. P. Gollub, 1988: 
Chaotic particle transport in time-dependent Rayleigh-Benard convection.   
{\it  Phys. Rev. A}, {\bf 38}, 6280--6286. 

Sullivan, P. P., J. C.~McWilliams, and C.-H.~Moeng, 1994:
A subgrid-scale model for large-eddy simulation of planetary boundary layer
flows. {\it Bound. Layer Meteorol.}, {\bf 71}, 247--276.

Taylor, G., 1921: Diffusion by continuous movement, 
{\it Proc. London Math. Soc.},{\bf 20}, 196--212.

Thomson, D. J., 1987:  
Criteria for the selection of stochastic models of particle trajectories in turbulent flows. 
{\it J. Fluid. Mech.}, {\bf 180}, 529--556.

Thomson, D. J., and B.J. Devenish, 2005:  
Particle pair separation in kinematic simulations. 
{\it J. Fluid. Mech.}, {\bf 526}, 277--302.

Willis, G. E., and J. W. Deardorff, 1976:  
A laboratory model of diffusion into the convective planetary boundary layer.  
{\it Q. J. R. Met. Soc.}, {\bf 102}, 427--445.

------ , and ------ , 1981:  
A laboratory study of dispersion from a source in the middle of the convective mixed layer.  
{\it Atm. Env.}, {\bf 15}, 109--117.


\newpage


\centerline{FIGURE CAPTIONS}

\bigskip
\noindent
Figure 1: wave pattern and isolines of the basic $2D$ stream function $\Psi_I$ (or equivalently $\Psi_{II}$). 
If the wave pattern is steady, all particles follow the $\Psi-$isolines. In the DSF model, the convective 
structures formed by the interplay between $\Psi_{I}$ and $\Psi_{II}$ are three-dimensional. 
Plot in arbitrary units.  

\bigskip
\noindent
Figure 2: FSLE $\lambda(\delta)$, with $\beta=\sqrt{2}$, of the single mode DSF, at seven different wavelength $l_1$. 
The LLE ($\lambda$) is the value of the plateau level ($\lambda(\delta)=const.$) 
and scales as $\lambda \sim l_1^{-2/3}$. Standard diffusion corresponds to the $\delta^{-2}$ 
regime. The FSLE "knees" follow the Richardson's scaling $\delta^{-2/3}$. All curves are smoothed by means 
of cubic spline interpolation. Statistics over $5000$ particle pairs.  

\bigskip
\noindent
Figure 3: FSLE $\lambda(\delta)$, with $\beta=\sqrt{2}$, of the multiscale DSF model: 
($+$) with "sweeping effect" correction, ($\times$) 
without "sweeping effect" correction (see Tab. II for details about the parameter set-up). 
The $\sim \delta^{-2/3}$ scaling, appearing in the case of "sweeping effect" 
correction, corresponds to the Richardson's law;  
the $\sim \delta^{-1/3}$ and $\sim \delta^{-4/9}$ scalings represent the possible 
Thomson and Devenish (2005) predictions for the case with no "sweeping effect" correction.
Statistics over $5000$ particle pairs. 

\bigskip
\noindent
Figure 4: relative dispersion $R^2(t)$ of the multiscale DSF model. Left curve: with "sweeping effect" correction; 
right curve: without "sweeping effect" correction (see Tab. II for details about the parameter set-up). 
The $t^3$ Richardson's scaling is reduced by overlap effects 
at the boundaries of the inertial range. The $t^{9/2}$ and $t^6$ scalings (Thomson and Devenish, 2005) are plotted 
as possible predictions for the no "sweeping effect" case. 
Statistics over $5000$ particle pairs.

\bigskip
\noindent
Figure 5: Mean square Lagrangian velocity difference between two trajectories as function of their separation, 
for the multiscale DSF model (see Tab. III). 
The quantities $\Delta V_i$ ($ms^{-1}$) and $l_i$ ($m$) are, 
respectively, the $i$-th component of the velocity difference and the 
$i$-th component of the distance between two particles ($i=1,2,3$). The (maximal) error bars are obtained 
averaging over the three spatial components.  
The Kolmogorov scaling $\sim l_i^{2/3}$ is followed inside the inertial 
range of the model. Statistics over 5000 particle pairs.

\bigskip
\noindent
Figure 6: Mean square absolute dispersion ($m^2$), as function of time ($s$), along the three directions 
for the multiscale DSF model (see Tab. III). Standard diffusion 
is approached on a time scale of the order of the largest turnover time. Trajectory dispersion is shown 
to be isotropic to a good extent.  Statistic over 5000 trajectories.

\bigskip
\noindent
Figure 7: Second-order temporal structure function of the three Lagrangian velocity components, for the 
multiscale DSF model (see Tab. III). The curves confirm a low degree 
of anisotropy (no more than $10-20\%$) in the Lagrangian trajectory motion. Statistics over 5000 trajectories.   

\bigskip
\noindent
Figure 8: Four snapshots showing the 
spreading of an ensemble of $10^4$ particles, initially distributed along a horizontal line placed 
at half height inside a box of edges $L_x=L_y=L_z=l_1^{(1)}=400$ $m$, having turnover time $\simeq 700$ $s$, 
 of the multiscale DSF model (see Tab. III), at four instants of time. Top left: $t=100$ $s$; 
 top right: $t=200$ $s$; bottom left: $t=500$ $s$; bottom right $t=1000$ $s$. Particle coordinates 
 are plotted assuming periodic boundary conditions relatively to the edges of the 
 box.

\bigskip
\noindent
Figure 9: FSLE $\lambda(\delta)$, with $\beta=\sqrt{2}$, for LES only ($+$), and LES+DSF ($\times$). 
See Tables I and IV for details. 
The $\alpha\delta^{-2/3}$ scaling corresponds to the Richardson's law with 
$\alpha \simeq 10^{-1}$ ${\rm m^{2/3}s^{-1}}$. In the uncoupled case, the "knee" of the 
plateau ($\lambda(\delta)=const.$) corresponds approximately to twice the LES grid step, 
$\delta \simeq 2
\Delta L$. The upper bound of the LES inertial range lies at about $\delta \simeq 400$ $m$.   
Statistics over $4096$ particle pairs.

\bigskip
\noindent
Figure 10: relative dispersion $R^2(t)$ for LES only ($+$) and LES+DSF ($\times$). Time $t$ is in s. 
See Tables I and IV for details. In the LES+DSF case $R^2(t)$ approaches the $t^3$ Richardson's law better than 
in the case of no coupling. Statistics over $4096$ particle pairs. The standard diffusion regime begins after 
about $4\tau_* \simeq 2000$ $s$.

\bigskip
\noindent
Figure 11: absolute dispersion $S^2(t)$ for LES only ($\circ$), and LES+DSF ($+$). Time $t$ is in s. 
See Tables I and IV for details. The coupling with the DSF model does not affect the behavior of $S^2(t)$. 
The standard diffusion regime, $\sim t$, is approached after about $4\tau_* \simeq 2000$ $s$.  
Statistics over $2 \, \times \, 4096$ trajectories.

\newpage


\begin{table}[h]
\begin{tabular}{|c|c|c|c|c|c|c|c|c|}
\hline 
$(N_x,N_y,N_z)$ & $(L_x,L_y,L_z)$ & $\Delta L$ & $(U_g,V_g)$ & $Q_*$ & $(z_{i})$ & $\tau_*$ & 
$\epsilon_{LES}$ & $dt$\\
$$  &  ($km$)  & $(m)$ & ($ms^{-1}$)  &  (${ms^{-1}K}$) & ($m$) & ($s$)  
 & ($m^2s^{-3}$) & ($s$)\\
\hline 
 $(128,128,128)$ & $(5,5,2)$ & $\simeq 39$ & $(10,0)$ & $0.24$  & $1000$  & $550$ & 
 $\simeq 10^{-3}$ & $1$\\
\hline
\end{tabular}
\caption{Basic parameters of the convective PBL flow simulated by LES. From left to right: grid points, 
domain size, grid step, horizontal geostrophic wind, kinematic heat flux, 
initial inversion height, turnover time, mean energy flux, integration time step.}
\end{table}

\newpage



\begin{table}[h]
\begin{tabular}{|c|c|c|c|c|c|c|c|c|c|c|c|c|}
\hline 
$N_m$ & $\epsilon_{DSF}$ & $l_1^{(1)}$ & $l_1^{(N_m)}$ & $A^{(1)}$ & $\tau^{(N_m)}$ & $A_{max}$ & $A_{rms}$
& $V_d$ & $dt$ & $\frac{dt}{\tau_{min}}$ & $\frac{dt}{\tau_{typ}}$ & 
$\frac{dt}{\tau^{(N_m)}}$\\
 $$ &  ($m^2s^{-3}$) & ($m$) & ($m$) & ($ms^{-1}$) & ($s$) & ($ms^{-1}$) & ($ms^{-1}$) & ($ms^{-1}$) 
 & ($s$) & $$ & $$ & $$\\
\hline 
$31$ & $10^{-3}$ & $40$ & $0.22$ & $0.13$ & $9.5$ & $1.95$ & $0.07$ & $1$ & $10^{-2}$ & 
$0.27$ & $10^{-1}$ & $10^{-3}$ \\
\hline
\end{tabular}
\caption{Set-up of the DSF model with external drift (Figs. \ref{fig:fsle_km} and 
\ref{fig:rd_km}), with $\rho=2^{1/4}$ and $C_K=0.25$. The eddy turnover times are  
$\tau^{(n)}=l_1^{(n)} / A^{(n)}$. 
The shortest advective 
time and the typical advective time across the smallest eddies are defined as, respectively, 
$\tau_{min}=l_1^{(N_m)}/2(A_{max}+V_d)$ and 
$\tau_{typ} =l_1^{(N_m)}/2(A_{rms}+V_d)$, where $A_{max}=\sum_{n=1}^{N_m} A^{(n)}$  and 
$A_{rms} =  \left\{ N_m^{-1} \sum_{n=1}^{N_m} \left[ A^{(n)} \right]^2 \right\}^{1/2}$. 
Amplitudes $\xi_i^{(n)}$ and pulsations $\omega_i^{(n)}$
the $n$ mode time oscillating perturbative terms are defined as in Tab. IV.}
\end{table}

\newpage

\begin{table}[h]
\begin{tabular}{|c|c|c|c|c|c|c|c|c|c|c|c|c|}
\hline 
$N_m$ & $\epsilon_{DSF}$ & $l_1^{(1)}$ & $l_1^{(N_m)}$ & $A^{(1)}$ & $\tau^{(N_m)}$ & $A_{max}$ & $A_{rms}$
& $V_d$ & $dt$ & $\frac{dt}{\tau_{min}}$ & $\frac{dt}{\tau_{typ}}$ & 
$\frac{dt}{\tau^{(N_m)}}$\\
 $$ &  ($m^2s^{-3}$) & ($m$) & ($m$) & ($ms^{-1}$) & ($s$) & ($ms^{-1}$) & ($ms^{-1}$) & ($ms^{-1}$) 
 & ($s$) & $$ & $$ & $$\\
\hline 
$45$ & $8 \cdot 10^{-3}$ & $400$ & $0.2$ & $0.56$ & $4.4$ & $9.3$ & $0.25$ & $0$ & $10^{-2}$ & 
$0.95$ & $10^{-2}$ & $10^{-3}$ \\
\hline
\end{tabular}
\caption{Set-up of the DSF model without external drift (Figs. \ref{fig:lsf_km},  
\ref{fig:ad_km}, \ref{fig:sf2_km} and \ref{fig:mixing_km}), with $\rho=2^{1/4}$ and $C_K=0.25$. 
The eddy turnover times are $\tau^{(n)}=l_1^{(n)} / A^{(n)}$. 
The quantities $\tau_{min}$, $\tau_{typ}$, $A_{max}$ and $A_{rms}$ (with $V_d=0$) are defined as in Tab. II.  
Amplitudes $\xi_i^{(n)}$ and pulsations $\omega_i^{(n)}$ 
the $n$ mode time oscillating perturbative terms are defined as in Tab. IV.}
\end{table}

\newpage

\begin{table}[h]
\begin{tabular}{|c|c|c|c|c|c|c|c|}
\hline 
$N_m$ & $l_1^{(1)}$ & $l_1^{(N_m)}$ & $A^{(1)}$ & $\tau^{(N_m)}$ & 
 $\frac{\xi_i^{(n)}}{\l_i^{(n)}}$ & $\omega_i^{(n)} \cdot \tau^{(n)}$ & $dt$\\
 $$ & ($m$) & ($m$) & ($ms^{-1}$) & ($s$) & $$ & $$  & ($s$)\\
\hline 
$43$ & $312$ & $0.2$ & $0.24$ & $10$ & $0.25$ & $\simeq 2\pi$  & $1$\\ 
\hline
\end{tabular}
\caption{Set-up of the DSF model as subgrid field of the convective LES 
(Figs. \ref{fig:fsle_lesdsf}, \ref{fig:rd_lesdsf} and \ref{fig:ad_lesdsf}), 
with $\rho=2^{1/4}$, $C_K=0.125$, $\epsilon_{DSF}=\epsilon_{LES}$ and $l_1^{(1)}=8 \Delta L$. 
The eddy turnover times are  $\tau^{(n)}=l_1^{(n)} / A^{(n)}$.  
$\xi_i^{(n)}$ and $\omega_i^{(n)}$, with $i=1,2,3$, are amplitudes and pulsations of the $n$ mode 
time oscillating perturbative terms.}
\end{table}

\newpage


\newpage


\end{document}